\pdfoutput=1
\documentclass[11pt,letterpaper]{article}

\newif\ifdraftmode
\draftmodetrue

\usepackage[utf8]{inputenc}
\usepackage[T1]{fontenc}
\usepackage{textcomp}
\usepackage{amsmath,amsthm}      
\usepackage{newtxtext}           
\usepackage{newtxmath}           
\usepackage{bm}                  
\usepackage{microtype}

\usepackage[letterpaper,margin=1in]{geometry}   
\usepackage[font=small,labelfont=bf,labelsep=period]{caption}
\usepackage{subcaption}
\usepackage{placeins}            
\usepackage{enumitem}
\setlist{itemsep=2pt,topsep=4pt,parsep=0pt}

\usepackage{graphicx}
\graphicspath{{figures/}}
\usepackage{xcolor}
\usepackage{tikz}                
\usetikzlibrary{arrows.meta,calc,fit,positioning} 
\usepackage{booktabs}            
\usepackage{multirow}
\usepackage{threeparttable}      
\usepackage{array}
\usepackage{siunitx}
\usepackage{algorithm}
\usepackage{algpseudocode}

\usepackage{xspace}
\usepackage{etoolbox}

\usepackage[round,authoryear,sort]{natbib}
\usepackage{authblk}

\usepackage[
  colorlinks=true,
  linkcolor={red!55!black},
  citecolor={blue!55!black},
  urlcolor={blue!55!black},
  breaklinks=true,
]{hyperref}
\usepackage[capitalize,nameinlink,noabbrev]{cleveref}
\crefname{appendix}{Appendix}{Appendices}
\Crefname{appendix}{Appendix}{Appendices}

\theoremstyle{definition}

\theoremstyle{remark}

\newcommand{\sidllm}{SID-LLM\xspace}
\newcommand{\sid}{SID\xspace}
\newcommand{\sids}{SIDs\xspace}
\newcommand{\rqkmeans}{RQ-KMeans\xspace}
\newcommand{\rqvae}{RQ-VAE\xspace}

\newcommand{\onedcg}[1]{OneDCG@#1}

\newcommand{\recall}[1]{Recall@#1}

\newcommand{\E}{\mathbb{E}}

\definecolor{ink}{HTML}{0B0B0B}
\definecolor{inksecondary}{HTML}{52514E}
\definecolor{muted}{HTML}{898781}
\definecolor{grid}{HTML}{E1E0D9}
\definecolor{axis}{HTML}{C3C2B7}
\definecolor{extoken}{HTML}{4A3AA7}
\definecolor{exsid}{HTML}{2A78D6}
\definecolor{exloss}{HTML}{EB6834}
\newcommand{\extok}[1]{{\color{extoken}\detokenize{<|#1|>}}}
\newcommand{\exsid}{{\color{exsid}\detokenize{<|sid_start|>...<|sid_end|>}}}
\newcommand{\exevent}[2]{\extok{dt_#1}\extok{#2}\exsid}
\newcommand{\exelided}{{\color{muted}...}}
\newcommand{\excols}[1]{\hspace*{#1\fontcharwd\font`\x}}
\newcommand{\exvdots}{\excols{2}\tikz[baseline=0pt]{\foreach \y in {0.5pt,3pt,5.5pt} \fill[fill=muted] (0,\y) circle (0.55pt);}}
\newcommand{\exloss}[1]{\sbox0{#1}\rlap{\hspace*{-0.8pt}\smash{\color{exloss!13}\rule[-2.3pt]{\dimexpr\wd0+1.6pt\relax}{8.4pt}}}\usebox0}
\newlength{\excardwidth}
\newcommand{\exsep}{\par\nointerlineskip\vspace{4pt}\noindent
  \tikz\draw[draw=axis, line width=0.4pt, dash pattern=on 2pt off 1.5pt] (0,0) -- (\excardwidth,0);\par\nointerlineskip\vspace{3pt}}
\newcommand{\excard}[3]{%
  \begin{tikzpicture}[baseline=(current bounding box.north)]
    \node[draw=axis, line width=0.5pt, rounded corners=3pt, inner sep=4pt, anchor=north west,
      text width=\excardwidth, align=left] (c) {%
      \ttfamily\fontsize{7.5}{9}\selectfont\color{ink}\raggedright #2%
      \if\relax\detokenize{#3}\relax\par
      \else\par\nointerlineskip\vspace{3pt}%
      {\setlength{\fboxsep}{2.5pt}\noindent\hspace*{-\fboxsep}\colorbox{exloss!13}{\begin{minipage}[t]{\excardwidth}%
        \raggedright #3\end{minipage}}\hspace*{-\fboxsep}}\fi};
    \node[anchor=south west, fill=grid, draw=axis, line width=0.5pt, rounded corners=3pt,
      font=\footnotesize\bfseries, inner xsep=4pt, inner ysep=1.5pt, minimum width=\excardwidth+8pt] at (c.north west) {#1};
  \end{tikzpicture}%
  \par\vspace{5pt}}

\newcommand{\todo}[1]{\ifdraftmode{\color{magenta}\sffamily\bfseries[TODO: #1]}\fi}

\definecolor{placeholder}{HTML}{C8500A}

\title{Generative Search and Recommendation at Shopify}

\author{%
  Juhana Kangaspunta\textsuperscript{*},
  Khalil Hajji\textsuperscript{*},
  Darshan Fofadiya\textsuperscript{*},
  Kyumin Park,
  Michael Garner,
  Anthony Maltsev\textsuperscript{\textdagger},
  Ryan Synk,
  Jose Sepulveda,
  Joshua Crisologo,
  Shuying Sun\textsuperscript{\textdaggerdbl},
  Akshay Soni\textsuperscript{\textdaggerdbl}
  \\
  Shopify
}

\date{}

\hypersetup{
  pdftitle={Generative Search and Recommendation at Shopify},
  pdfauthor={Shopify},
  pdfkeywords={semantic IDs, generative retrieval, large language models, recommendation, search, agentic commerce},
}

\begin{document}

\maketitle

\begin{NoHyper}
  \renewcommand{\thefootnote}{}
  \footnotetext{%
    \textsuperscript{*}Equal contribution.\quad
    \textsuperscript{\textdagger}Liquid AI.\quad
    \textsuperscript{\textdaggerdbl}Project supervision.
  }
\end{NoHyper}

\vspace{-1.5em}

\begin{abstract}
Search and recommendation at Shopify require retrieving relevant products from a changing catalog of billions of products under strict latency constraints. We present \sidllm, which casts both tasks as generating semantic IDs: short sequences of content-derived tokens representing products. We jointly pre-train \sidllm on catalog and behavioral data, giving search and recommendation a shared representation and training objective. We show that reinforcement learning further improves judged search relevance. Furthermore, distillation from task-specific experts improves recommendation quality over joint supervised fine-tuning without a statistically detected loss in search quality. Finally, for efficient serving, we pair multi-token prediction with a lightweight ranker, improving retrieval quality over our autoregressive baseline while reducing latency.

We evaluate \sidllm both as a standalone retriever and as a component of Shopify's production search and recommendation systems. As a standalone search system, \sidllm achieves quality comparable to Shopify's existing multi-model search stack while meeting production latency requirements. Integrating \sidllm into the search and recommendation stacks provides significant gains in both offline evaluations and online A/B tests. Together, these results demonstrate the practical value of semantic-ID generation as a unified approach to product discovery at scale.
\end{abstract}
\section{Introduction}
\label{sec:introduction}
Search and recommendation connect buyers with relevant products across Shopify’s large and diverse catalog across millions of merchants on their web storefront, Shopify-built Shop App, and Shopify's Catalog API designed for conversational commerce and AI agents. Both require understanding products and their relevance to a particular buyer at the scale of billions of products with very low latency.

We present \sidllm, a generative model for search and recommendation. It represents
products as semantic IDs (\sids), short sequences of content-derived tokens, and learns to
generate relevant identifiers from the context available for each task. This gives search
and recommendation a shared formulation and allows us to investigate unifying them within
one model.

This report describes \sidllm from semantic-ID tokenization and training through production
integration. We study the choices that determine retrieval quality and serving
cost, and explore distillation from task experts into a unified student. Adding \sidllm
provides significant gains in both online and offline evaluations of Shopify's search
and recommendation stacks while meeting strict latency requirements for online serving.

\begin{figure}[!t]
  \centering
  \includegraphics[width=0.90\linewidth]{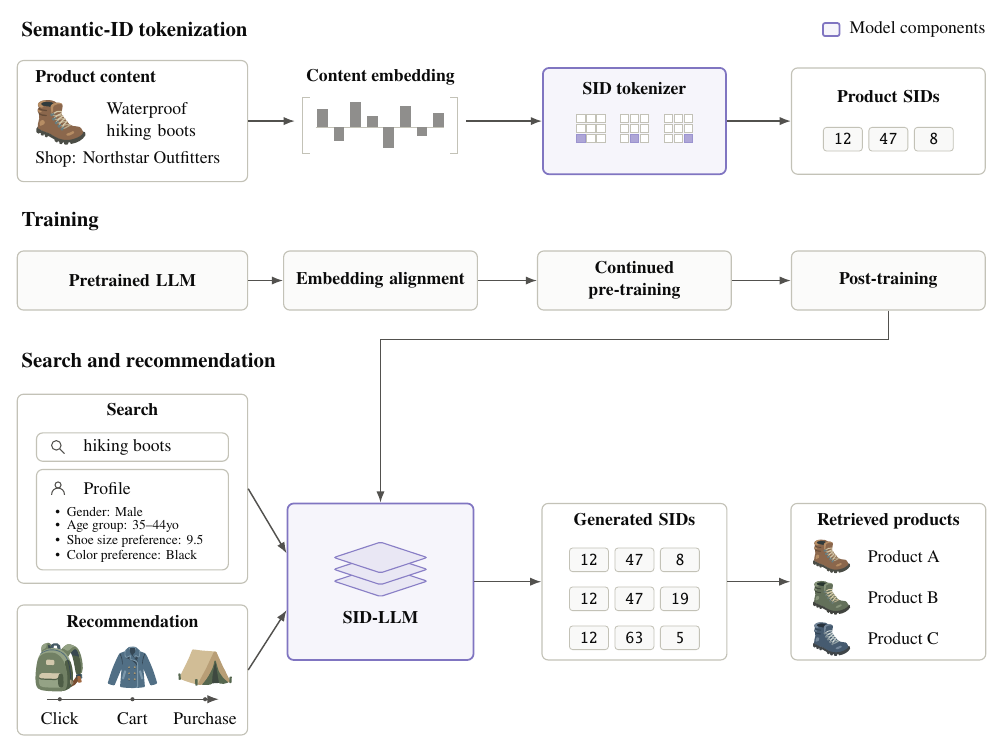}
  \caption{Overview of \sidllm.}
  \label{fig:sid-llm-overview}
\end{figure}
\FloatBarrier

\subsection{Task formulations}
\label{sec:intro-tasks}
Shopify serves product search through the Catalog API, merchant storefronts and the Shop app.
Recommendation surfaces include the Shop app's home feed and ads. These experiences operate
over a multi-merchant catalog, with product eligibility depending on the request context,
such as the buyer's market.

Each task retrieves candidate products for downstream ranking
(\cref{sec:integration}). We distinguish three tasks according to the context available
at retrieval time.

\paragraph{Recommendation.}
Given a buyer's interaction history $h$, retrieve products they are likely to find useful
or purchase. The history contains actions such as clicks, carts, favorites and purchases,
together with the products involved and their timing. There is no explicit query expressing
the buyer's current need, so the model must infer it from this context. The task is to
identify products the buyer is likely to engage with next.

\paragraph{Non-personalized search.}
Given a textual query $q$, retrieve products relevant to the expressed intent without
conditioning on buyer-specific preferences. Relevance requires interpreting the requested
product type, attributes and constraints even when the buyer and merchant describe the
same product differently.

\paragraph{Personalized search.}
Given a query $q$ and a buyer profile $b$, retrieve products that satisfy the query while
accounting for the buyer's preferences. The profile summarizes information about the buyer
derived from their past behavior. The query specifies the immediate need; the profile helps
distinguish among products that could meet it.

\paragraph{Retrieval as semantic-ID generation.}
We represent each product by a short semantic ID obtained by quantizing its content embedding
(\cref{sec:tokenizer}). Given the task-specific context, \sidllm learns to generate target
\sids through next-token prediction; we also study predicting their tokens in parallel
(\cref{sec:sid-llm-mtp}). At inference, constrained decoding produces candidate \sids,
which are resolved to eligible products before downstream ranking (\cref{sec:integration})

\subsection{Traditional recommendation systems vs.\ \sidllm}
\label{sec:intro-traditional}
Industrial recommendation systems commonly separate candidate retrieval from ranking
\citep{covington2016youtube}. A two-tower retriever \citep{yi2019twotower} encodes the buyer
and each product separately and scores their compatibility with an inner product. This
supports efficient nearest-neighbour retrieval, but compresses the buyer's history into
a fixed vector before scoring candidates. Multi-interest methods \citep{li2019mind} use
several buyer vectors, each still computed independently of the candidate.

Sequential generative recommenders learn to predict the next item from the interaction
sequence \citep{zhai2024hstu}. Representing products as atomic IDs gives each product its own
learned embedding, with no content-based relationship between identifiers. Semantic-ID
generation \citep{rajput2023tiger} instead shares tokens across related products. In \sidllm,
the history remains part of the prompt throughout generation, and a pretrained language
backbone provides an initial representation of product language. We evaluate the resulting
generator both on its own candidates and on the purchases it adds to the combined candidate
pool (\cref{sec:eval-rec-pool}).

\subsection{Lexical matching, semantic search, and generative search}
\label{sec:intro-paradigms}
Lexical retrieval matches query terms to product text through an inverted index, using a
scoring function such as BM25 \citep{robertson2009bm25}. It is useful for exact terms such as
brands and model numbers, but encounters a vocabulary mismatch when buyers and merchants
use different words for the same thing \citep{furnas1987vocabulary}. Dense semantic retrieval
addresses this mismatch by encoding queries and products into a shared vector space
\citep{karpukhin2020dpr,nigam2019semantic} and searching an approximate nearest-neighbour
index \citep{jegou2011pq,malkov2020hnsw}. Like two-tower recommendation, it scores separately
computed representations.

Generative retrieval replaces the index lookup with identifier generation
\citep{tay2022dsi,decao2021genre}. In \sidllm, the query and optional buyer profile condition
the prediction of each \sid token. The model learns which identifiers fit the context,
while a catalog index constrains generation to identifiers that resolve to eligible products
(\cref{sec:inference-artifacts}). Shopify's search stack combines lexical, embedding and
generative retrieval, and we evaluate \sidllm's contribution through end-to-end stack ablations
(\cref{sec:eval-search-offline}).

\paragraph{Contributions.}
\begin{itemize}
  \item A study of semantic-ID tokenization, including quantization methods, codebook sizes
    and product collisions (\cref{sec:tokenizer}).
  \item A comparison of language backbones and autoregressive versus multi-token prediction
    for generative retrieval (\cref{sec:sid-llm}).
  \item A training recipe combining catalog, search and recommendation data, with ablations
    of the data mixture, post-training objectives and distillation from task experts
    (\cref{sec:pretraining,sec:post-training}).
  \item Production integration and inference methods for fast, eligibility-constrained
    retrieval over a changing catalog (\cref{sec:integration,sec:inference}).
  \item Offline and online evaluation of the gains from adding \sidllm to the search and
    recommendation stacks (\cref{sec:evaluation}).
\end{itemize}
\section{Evaluation Metrics}
\label{sec:eval-metrics}
We use three primary offline metrics: search relevance measured by \onedcg{10}, search purchase
\recall{10}, and recommendation \recall{50}. We report additional metrics only for specific analyses,
such as discovery and incremental candidate-pool recall.

\label{sec:eval-protocols}
Offline scores depend on more than the model: the relevance judge (\cref{sec:eval-ndcg}), the query set, the
retrieval index, the users scored for recommendation and the inference code all changed during the project. We call one fixed
combination an evaluation protocol (\cref{tab:eval-protocols} in the appendix). Tables using these
protocols name them in their captions; separate recommendation cohorts are described alongside their
results. Absolute scores compare only within a table.

\subsection{Search}

\subsubsection{OneDCG}
\label{sec:eval-ndcg}
An LLM judge grades each returned product from 0 to 3 for its relevance to the query and to the
profile of the buyer who searched. The grade $r$ and a static score $s$, which depends on neither
the query nor the buyer, set the product's gain, $100\,(2^{r+s}-1)/15$. The static score is a
weighted sum of four quality signals between 0 and 1: the product's ratings and popularity, the
brand's prevalence and the shop's sales. Its weights sum to one, so the gain runs from 0 to 100 and
the static score is worth at most one grade. A perfectly relevant product with every quality signal
at zero gains 46.7. Protocols E1 and E2 (\cref{tab:eval-protocols}) use earlier versions of the
judge, which read the query alone and combine the grade with the quality signals differently. We
report \onedcg{$k$}: the DCG@$k$ \citep{jarvelin2002ndcg} of these gains, divided by the DCG of $k$
results that all receive the maximum gain of 100. Unlike in NDCG, the normalizer is a constant rather than each
query's ideal DCG, so the score measures absolute relevance and is comparable across queries.

\subsubsection{Purchase recall}
\label{sec:eval-purchase-recall}
Purchase \recall{$k$} is the share of a converted search session's purchased products that
appear among the first $k$ retrieved, averaged over sessions. Products are matched across shops
by a universal product identifier, and a purchase without one counts as a miss.

\subsection{Recommendation}

\subsubsection{Recommendation recall}
\label{sec:eval-rec-recall}
Recommendation \recall{$k$} scores the products a user buys in a held-out week
after their interaction history: the share of them among the first $k$ recommended, averaged
over the users with at least one such purchase.

\section{Semantic ID Tokenizer Model}
\label{sec:tokenizer}
The tokenizer maps each product to a short semantic ID (\sid), which the downstream LLM uses
as a sequence of discrete tokens (\cref{sec:sid-llm}). We write codebook sizes in level order,
separated by hyphens. Our final tokenizer uses residual $k$-means (\rqkmeans) with codebooks
\mbox{8192-8192-8192}, meaning three levels of 8,192 codes each.

\subsection{Tokenizer construction}
\label{sec:tokenizer-data}
\label{sec:tokenizer-training}
We start from one 768-dimensional embedding per product, produced by an in-house model trained
for product search. \rqkmeans quantizes the embedding in stages: the first codebook approximates
the original vector, and each subsequent codebook approximates the residual left by the previous
levels. The three centroid indices form the product's \sid.

We fit on the 25M products with the highest gross merchandise value (GMV), a sample that gives
higher downstream search relevance than random sampling in both codebook shapes tested
(\cref{tab:fit-sample} in the appendix). We fit each level with standard $k$-means on GPU
(cuML), using scalable $k$-means++ initialization and up to 20 iterations with early stopping
at convergence. Once the codebooks are fitted, we assign every product greedily to the nearest
centroid at each level.

\subsection{Design choices}
\label{sec:tokenizer-model}
\label{sec:tokenizer-eval}
We select tokenizers using downstream retrieval quality and serving latency. Each comparison
trains an LLM on the candidate \sids and scores it against a baseline under the same evaluation
protocol (\cref{sec:eval-protocols}). We screen with small models and short runs, then confirm
promising choices with larger or longer runs and load-test the finalists.

\paragraph{Quantization method.}
\rqvae \citep{lee2022rqvae} learns a neural encoder before residual quantization; \rqkmeans
quantizes the product embeddings directly. At matched codebook shape, \rqkmeans gives higher
\onedcg{10} despite producing fewer unique \sids (\cref{tab:rq-vae-vs-rq-kmeans}). This motivates
our choice of \rqkmeans and shows that uniqueness alone is insufficient for selecting a tokenizer.
Similar findings have been reported in industry \citep{damico2026glide,zhou2025openonerec}.

\begin{table}[!htbp]
  \caption{\rqvae vs.\ \rqkmeans as the tokenizer of a downstream LLM. All codebooks are fit on the same 
    sample of product embeddings. Unique SIDs:
    distinct SIDs as a share of all products. Resnik: mean similarity of purchased-item pairs
    for the same query, computed separately from August search logs (\cref{sec:appendix-resnik}).
    \onedcg{10}: evaluation protocol E1 (\cref{tab:eval-protocols}).
    Best \onedcg{10} in bold.}
  \label{tab:rq-vae-vs-rq-kmeans}
  \centering\footnotesize
\begin{tabular}{l l S[table-format=2.1] S[table-format=1.2] S[table-format=1.4]}
\toprule
Method & Codebooks & {Unique SIDs (\%)} & {Resnik (bits)} & {\onedcg{10}} \\
\midrule
\rqvae & 1024-1024-1024-1024 & 69.1 & 6.25 & 0.3518 \\
\rqkmeans & 1024-1024-1024-1024 & 35.3 & 8.32 & 0.3804 \\
\midrule
\rqkmeans & 8192-8192-8192 & 30.4 & 8.68 & \bfseries 0.3945 \\
\bottomrule
\end{tabular}

\end{table}

\paragraph{Behavioral coherence.}
To examine how well the tokenizers group products that satisfy similar search intents, we adapt
Resnik similarity \citep{resnik1995information} to the \sid prefix tree. We score pairs of distinct
items purchased for the same query using the information content of their longest shared \sid
prefix: sharing a prefix that covers fewer items receives a higher score. We average over item
pairs within each query, then equally across 12,799 queries. Across these three configurations,
higher similarity accompanies higher downstream search quality, despite lower \sid uniqueness
(\cref{tab:rq-vae-vs-rq-kmeans}). This suggests that grouping products around shared search intent
is an important property of the representation. The data and metric are detailed in
\cref{sec:appendix-resnik}.

\paragraph{Codebook size.}
Larger codebooks improve retrieval quality but increase serving cost. The
\mbox{32768-8192-8192} tokenizer scores higher than \mbox{8192-8192-8192}, but its p95 latency
reaches 50\,ms at the target load of 100 requests per second (\cref{tab:codebook-latency}). We
choose \mbox{8192-8192-8192} to preserve latency headroom. The full size and balancing sweep and the
measurements at other request rates are in \cref{sec:appendix-tokenizer}.

\begin{table}[t]
  \caption{Codebook quality and p95 serving latency at 100 requests per second. Both \rqkmeans
    tokenizers are fit on the top 25M products by GMV, each with its own LFM2-2.6B downstream LLM,
    on one NVIDIA B200 node. \onedcg{10}: evaluation protocol E2
    (\cref{tab:eval-protocols}). Better value per column in bold.}
  \label{tab:codebook-latency}
  \centering\small
\begin{tabular}{l S[table-format=1.4] S[table-format=2.1]}
\toprule
Codebooks & {\onedcg{10}} & {p95 latency (ms)} \\
\midrule
8192-8192-8192 & 0.3464 & \bfseries 40.5 \\
32768-8192-8192 & \bfseries 0.3538 & 50.0 \\
\bottomrule
\end{tabular}

\end{table}

\subsection{Collisions and product resolution}
\label{sec:tokenizer-resolution}
Different products can share a \sid, including duplicate or closely related listings. Increasing
uniqueness through balancing or heuristic de-duplication does not consistently improve downstream
retrieval. Adding a fourth CRID token \citep{ling2026crid} gives a small relevance gain but
increases decoding latency, so we retain three semantic codewords. Detailed comparisons and the
cluster-size analysis are in \cref{sec:appendix-tokenizer}.

In search, we resolve each \sid to the eligible product with the highest static quality score
(\cref{sec:eval-ndcg}). Recommendation allows confident \sids to contribute multiple products,
using the joint-probability resolution described in \cref{sec:integration-rec}.
\FloatBarrier

\section{Semantic ID LLM Model}
\label{sec:sid-llm}

\subsection{Autoregressive vs.\ multi-token prediction}
\label{sec:sid-llm-mtp}
A \sid is a short, fixed-length sequence of codewords (three in our tokenizer, \cref{sec:tokenizer}),
and a \sidllm can emit it in two ways. \emph{Autoregressive} (AR) decoding predicts one codeword per
step: the prefill encodes the request, and each successive codeword is drawn from a distribution
conditioned on the codewords already chosen, so a length-three \sid costs one prefill and two further
backbone passes. \emph{Multi-token prediction} (MTP) instead predicts all three codeword distributions
from the request prefix alone, in a single prefill, with no codeword attending to another; the catalog
index constrains these independent per-position distributions to valid combinations, and a beam walk
over it assembles the candidate \sids. MTP trades the exact factorization of the joint \sid distribution
for parallelism---dropping the sequential dependence between codewords is precisely what lets it decode
in one pass instead of three.

The two modes expose a quality--latency trade-off (\cref{tab:ar-mtp}). On its own, MTP scores below AR
on both top-10 search metrics, and widening its beam from 100 to 250 barely moves either, so
more beam search does not close the gap. What MTP buys is speed: predicting the whole \sid in one prefill roughly halves median server-side latency, and its median
and p95 stay nearly flat as offered load rises while its p99 grows only to about 61\,ms at the highest
load---far below AR's, whose per-codeword passes make its tail balloon under load (\cref{tab:inference-latency}).
Re-scoring
MTP's 250 candidates with the learned ranker (\cref{sec:inference-ranker}) more than recovers the gap,
reaching \onedcg{10} and \recall{10} above AR on both, while staying faster than AR at
every percentile. We therefore serve MTP paired with the ranker for search. Recommendation uses an AR
model with joint-probability product resolution (\cref{sec:integration-rec}). The unified student studied
in \cref{sec:post-training-unification} is experimental and starts from the AR model.

\begin{table}[htbp]
  \caption{Autoregressive vs.\ multi-token prediction: LFM2-2.6B pre-trained to predict one \sid token
    per step (autoregressive) or the whole \sid at once (multi-token prediction). The multi-token model
    is decoded at two beam widths, then with a learned ranker that re-scores its 250 candidates. Every
    arm returns 100 products per query. Evaluation protocol E5 (\cref{tab:eval-protocols}). Best value
    per column in bold.}
  \label{tab:ar-mtp}
  \centering\footnotesize
\begin{tabular}{l S[table-format=3] S[table-format=1.4] S[table-format=1.4]}
\toprule
 &  & \multicolumn{1}{c}{Search relevance} & \multicolumn{1}{c}{Search purchases} \\
\cmidrule(lr){3-3} \cmidrule(lr){4-4}
Arm & {Beam width} & {\onedcg{10}} & {\recall{10}} \\
\midrule
Autoregressive & 100 & 0.3083 & 0.3246 \\
Multi-token prediction & 100 & 0.2804 & 0.2839 \\
Multi-token prediction & 250 & 0.2811 & 0.2840 \\
Multi-token prediction + ranker & 250 & \bfseries 0.3365 & \bfseries 0.3450 \\
\bottomrule
\end{tabular}

\end{table}

\subsection{Backbone ablations}
\label{sec:sid-llm-backbones}
We compare four backbones: LFM2-2.6B \citep{liquid2025lfm2}, Qwen3-1.7B-Base,
Qwen3-4B-Base and Qwen3-14B-Base \citep{yang2025qwen3}. Training recipes differ
across backbones (\cref{sec:pretraining-training,tab:training-recipe}); the 1.7B
model follows the 4B recipe. \Cref{tab:backbones} summarizes the quality--latency
trade-off. The Qwen3 models improve average quality at higher latency. We deploy
the LFM2-2.6B backbone, which has the lowest p95 latency in this comparison.

\begin{table}[htbp]
  \caption{Backbone quality--latency trade-off. Mean quality gain is the
    equal-weight mean relative gain over LFM2-2.6B in search relevance
    (\onedcg{10}), search purchases (\recall{10}) and recommendation
    (\recall{50}). Latency change is relative to the same baseline; positive
    values indicate higher latency. Evaluation protocol E5
    (\cref{tab:eval-protocols}).}
  \label{tab:backbones}
  \centering\small\setlength{\tabcolsep}{5pt}
  \begin{tabular}{lrrr}
    \toprule
    Backbone & p95 latency (ms) & Latency change (\%) & Mean quality gain (\%) \\
    \midrule
    LFM2-2.6B & 30.35 & 0.0 & 0.0 \\
    Qwen3-1.7B-Base & 73.29 & +141.5 & +1.0 \\
    Qwen3-4B-Base & 105.00 & +246.0 & +3.5 \\
    Qwen3-14B-Base & 142.87 & +370.7 & +4.9 \\
    \bottomrule
  \end{tabular}
\end{table}

\section{Pre-training}
\label{sec:pretraining}

\subsection{Data}
\label{sec:pretraining-data}
Our goal is to make \sidllm perform well on both the search and recommendation tasks. In order to do that, we pre-train the model with three different
categories of data: product attribute data, search data and recommendation data (\cref{tab:mixture}). Furthermore, we explore adding general text, math and coding replay
data during the pre-training phase and use that for the larger models (\cref{sec:pretraining-replay}).

During training, we split each training example into a prompt and a completion, and apply the loss only to the completion. The details on how each of
the examples is split can be seen in \cref{fig:training-examples} and in the following sections.

In order to evaluate the importance and appropriate dosage of the different data streams, we perform shorter training runs ($\sim$30k steps) with different data mix configurations.

\begin{table}[htbp]
  \caption{Pre-training data mix of LFM2-2.6B: 100,000 steps of 1,048,576 tokens each. Epochs are passes over a stream's
    rows. ``Loss tokens'' counts only those that carry the loss; ``Tokens'' counts all.}
  \label{tab:mixture}
  \centering\footnotesize
\begin{tabular}{l S[table-format=4.1] S[table-format=2.2] S[table-format=3.2] S[table-format=3.1] S[table-format=3.1]}
\toprule
Source & {Examples (M)} & {Epochs} & {Tokens (B)} & {Tokens (\%)} & {Loss tokens (\%)} \\
\midrule
\multicolumn{6}{l}{\itshape Product attributes} \\
\quad Product documents & 913.0 & 1.33 & 32.41 & 30.9 & 31.0 \\
\quad Top-GMV product documents & 226.2 & 8.00 & 6.76 & 6.5 & 7.7 \\
\addlinespace[3pt]
\multicolumn{6}{l}{\itshape Search} \\
\quad Search session examples & 144.5 & 3.88 & 41.46 & 39.5 & 5.3 \\
\quad LLM-judged examples & 61.9 & 3.00 & 1.41 & 1.3 & 1.7 \\
\quad LLM-judged personalized examples & 1.1 & 76.20 & 0.65 & 0.6 & 0.2 \\
\addlinespace[3pt]
\multicolumn{6}{l}{\itshape Recommendation} \\
\quad Interaction sequences & 34.9 & 0.10 & 12.32 & 11.8 & 27.1 \\
\quad New-product purchases & 57.1 & 0.23 & 5.67 & 5.4 & 16.3 \\
\quad New-merchant purchases & 50.3 & 0.25 & 3.12 & 3.0 & 8.1 \\
\quad New-category purchases & 14.2 & 0.06 & 1.04 & 1.0 & 2.8 \\
\midrule
Total & 1503.2 & {--} & 104.86 & 100.0 & 100.0 \\
\bottomrule
\end{tabular}

\end{table}

\begin{figure}[htbp]
  \setlength{\excardwidth}{\dimexpr 3.17in-8pt\relax}%
  \noindent\begin{minipage}[t]{3.17in}
  \excard{Product attributes}
    {Attribute to product id mapping:\\
     - simplified\_title: waterproof trail running shoe\\
     - id: \exloss{\exsid}\exsep
     Attribute to product id mapping:\\
     - category: Apparel \& Accessories > Shoes >\\
     \excols{12}Athletic Shoes\\
     - id: \exloss{\exsid}}
    {}
  \excard{Search}
    {Personalized search:\\[2pt]
     User profile:\\
     \{"personas": [\{\\ 
     \excols{3}"interests": [\{"name": \exelided, "last\_seen": \exelided,\\ 
     \excols{18}"recommended\_products": [\exelided],\\ 
     \excols{18}"key\_brands": [\exelided]\}, \exelided],\\ 
     \excols{3}"summary": \exelided\}],\\ 
     \excols{1}"taste": \{"aesthetic": [\exelided],\\ 
     \excols{11}"color\_preferences": [\exelided]\}\}\\[2pt] 
     \extok{search_query} trail running shoes}
    {\extok{click}\exsid\\
     \extok{purchase}\exsid}
  \end{minipage}\hfill
  \begin{minipage}[t]{3.17in}
  \excard{Recommendation (CPT)}
    {User interactions:\\
     \exevent{>365d}{purchase}\\
     \exevent{90d}{cart}\\
     \exvdots}
    {\exevent{90d}{click}\\
     \exvdots\\
     \exevent{1d}{cart}\\
     \exevent{1d}{purchase}}
  \excard{Recommendation (SFT)}
    {User interactions:\\
     \exevent{>365d}{purchase}\\
     \exevent{90d}{cart}\\
     \exvdots\\
     \exevent{1d}{cart}\\
     \extok{dt_1d}\exloss{\exsid}}
    {}
  \end{minipage}\par\vspace{1pt}
  {\centering\scriptsize\color{inksecondary}
   {\ttfamily\extok{click}} special token\quad
   {\ttfamily\exsid} a product's \sid\quad
   \tikz[baseline=-0.6ex]\fill[fill=exloss!13] (0,-0.07) rectangle (0.3,0.07); completion (loss)\quad
   {\ttfamily\extok{dt_30d}} time since the previous event\par}
  \caption{Training examples of each kind of data. The shaded completion carries the loss; the prompt before it is
    masked. A product attribute example names one attribute. A recommendation completion holds the user's last
    30 days of events in the continued pre-training (CPT) format and one purchased product in the supervised
    fine-tuning (SFT) format.}
  \label{fig:training-examples}
\end{figure}

\subsubsection{Product attribute data}
\label{sec:pretraining-attribute-data}
To teach the model the product catalog and how textual product attributes map to \sids, we construct
a product attribute dataset from the catalog data. During training, we sample a single product attribute from the
following list: title, category, brand, price, other product attributes, shop name, and shop country code. We then make the model predict the \sid as the completion
based on the product attribute in the prompt.

We create two separate product attribute datasets: one containing all products in the catalog and one consisting only of products with at least \$1,000 of sales over the last year. Both datasets form the training examples
the same way, but the second one samples the attribute only from the title and the shop name.

We ablate the usefulness of the product attribute datasets to the final tasks, and notice that product attribute data has a positive impact
on search metrics, but does not affect the recommendation metrics meaningfully (\cref{tab:product-attributes}).

\begin{table}[htbp]
  \caption{Product attribute data: LFM2-2.6B pre-trained on the baseline mix (\cref{tab:baseline-mix})
    and on the same mix without its product attribute data, whose tokens go to the other streams in
    proportion. Evaluation protocol E3 (\cref{tab:eval-protocols}). Better value per column in
    bold.}
  \label{tab:product-attributes}
  \centering\footnotesize
\begin{tabular}{l S[table-format=1.4] S[table-format=1.4] S[table-format=1.4]}
\toprule
 & \multicolumn{1}{c}{Search relevance} & \multicolumn{1}{c}{Search purchases} & \multicolumn{1}{c}{Recommendation} \\
\cmidrule(lr){2-2} \cmidrule(lr){3-3} \cmidrule(lr){4-4}
Arm & {\onedcg{10}} & {\recall{10}} & {\recall{50}} \\
\midrule
Baseline & \bfseries 0.2876 & \bfseries 0.3336 & 0.2096 \\
No product attribute data & 0.2818 & 0.3250 & \bfseries 0.2099 \\
\bottomrule
\end{tabular}

\end{table}

\subsubsection{Search data}
\label{sec:pretraining-search-data}
We train the model with search data constructed from two sources: real search session interactions and synthetic LLM judgments.
\paragraph{Search session examples.}
We collect search sessions from the different Shopify search surfaces and construct training examples from them. Each training example represents a single search session for a given user.
As seen in \cref{fig:training-examples}, an example contains a JSON buyer profile (if available), the search query, and the \sids of the products the user interacted with, each prefixed with an interaction token. The interaction token indicates
which interaction type (click, purchase, add-to-cart, etc.)\ happened for that given product in this search session. The buyer profile is an LLM-written summary of the buyer's interests, brands, and taste.
\paragraph{LLM-judged examples.}
In order to extend beyond data from real search sessions, we generate more data using the existing search stack and an LLM judge. To do that, we sample real user search queries and use the existing search
system to generate candidate products for them. We then use an LLM judge to rate the relevance of each query-product pair. Next, we combine the relevance score
with a static product quality score to come up with a combined score for each pair. Finally, we construct training examples similar to the search session examples by thresholding on the combined score. Products that exceed the minimum
score threshold get labeled as clicks, products that exceed a higher threshold get labeled as purchases, and products that fall below the minimum threshold get excluded from the training example.

Additionally, we construct similar LLM-judged personalized examples where the main differences from the non-personalized examples are including a buyer profile as context and using a personalized LLM judge like the one in \cref{sec:eval-ndcg} to label the query-profile-product triplet for
relevance.


\subsubsection{Recommendation data}
\label{sec:pretraining-rec-data}
The recommendation data teaches \sidllm to predict what a buyer will interact with next, given their history. It
has two parts: interaction sequences and discovery streams.

\paragraph{Interaction sequences.} A buyer's clicks, carts, favorites, checkouts and purchases in time order. Each
event is a time-since-previous token, an action token and the product's \sid. The last 30 days of events are the
completion, and everything before is the prompt (\cref{fig:training-examples}). Raw histories are dominated by old
clicks that no longer reflect what the buyer wants, so we compress them: we drop clicks older than 30 days and carts,
favorites and checkouts older than 90 days, and keep purchases at any age. We apply the same compression to the
fine-tuning prompts, so the model sees the same kind of history in both stages.

\paragraph{Discovery streams.} Most of what a buyer interacts with next is something they have already seen, so a
model trained only on interaction sequences learns to repeat the history. The more valuable recommendations are the
products the buyer does not know yet. We add three streams that list only a buyer's first purchases: of a new
product, from a new merchant, and in a new category. Each stream has its own task description. Every event in these
streams is a discovery, so every event carries the loss.

We measure discovery with discovery \recall{50}: recall over target purchases of products the buyer has never
interacted with. Compared with the same recipe without these changes, the fine-tuned model gains about 10\% in
discovery \recall{50} and about 2\% in overall \recall{50} (\cref{tab:rec-discovery}).

\begin{table}[htbp]
  \caption{Effect of the discovery recipe on Qwen3-14B-Base, both models fine-tuned: the base recipe vs.\ the same
    recipe with compressed interaction histories and the three discovery streams. 250k purchasers
    (February 2026 validation set), \rqkmeans \mbox{8192-8192-8192} \sids, beam 50. Discovery recall is over the 152,700
    buyers with a new-product target. Best value per column in bold.}
  \label{tab:rec-discovery}
  \centering\footnotesize
\begin{tabular}{l S[table-format=1.4] S[table-format=1.4]}
\toprule
 & \multicolumn{1}{c}{Recall} & \multicolumn{1}{c}{Discovery recall} \\
\cmidrule(lr){2-2} \cmidrule(lr){3-3}
Model & {\recall{50}} & {\recall{50}} \\
\midrule
Base recipe & 0.2880 & 0.0585 \\
+ compression and discovery streams & \bfseries 0.2945 & \bfseries 0.0646 \\
\bottomrule
\end{tabular}

\end{table}

\subsubsection{General-domain replay data}
\label{sec:pretraining-replay}
We investigate whether maintaining the pre-trained base models' general-domain capabilities helps on the search and recommendation tasks. We ablate the
presence vs.\ absence of general-domain replay data during pre-training and measure model performance both on the search and recommendation tasks, as well as
general-domain tasks.

Our replay data consists of the following datasets: educational web text from FineWeb-Edu \citep{penedo2024fineweb}, algorithmic code from
OpenCoder \citep{huang2025opencoder} and math from SwallowMath-v2 \citep{fujii2025swallowmath}. When adding replay data, we scale the other datasets down to 90\%, and therefore assign 10\% of the loss tokens to replay data.
The token proportions are detailed in \cref{tab:baseline-mix}.

We ablate the replay data on both LFM2-2.6B and Qwen3-14B-Base (\cref{tab:replay}). For the LFM2-2.6B model, adding replay data
hurts both search and recommendation performance. However, for the larger Qwen3-14B-Base model, adding replay data has a roughly neutral
effect on those tasks. For both models, MMLU \citep{hendrycks2021mmlu} drops sharply if replay data is not present, meaning that
the models suffer catastrophic forgetting of their general capabilities. Adding replay data recovers some of the capabilities in both models. As our final goal
is to perform well on the search and recommendation tasks, we choose to add replay data only when training the larger models.

\begin{table}[htbp]
  \caption{General-domain replay at 10\% of the loss tokens (\cref{tab:baseline-mix}). LFM2-2.6B trains
    for 30,000 steps and Qwen3-14B-Base for 20,000, on the same mixes, whose shares are solved for
    the LFM2 tokenizer. Search and recommendation: evaluation protocol E3
    (\cref{tab:eval-protocols}). MMLU: 5-shot accuracy; chance is 0.25. Backbone: the released
    model, before any \sid training. Compare arms within a model only: training length and
    recommendation decoding differ between the models. Better arm per model in bold.}
  \label{tab:replay}
  \centering\footnotesize
\begin{tabular}{l S[table-format=1.4] S[table-format=1.4] S[table-format=1.4] S[table-format=1.4]}
\toprule
 & \multicolumn{1}{c}{Search relevance} & \multicolumn{1}{c}{Search purchases} & \multicolumn{1}{c}{Recommendation} & \multicolumn{1}{c}{MMLU} \\
\cmidrule(lr){2-2} \cmidrule(lr){3-3} \cmidrule(lr){4-4} \cmidrule(lr){5-5}
Arm & {\onedcg{10}} & {\recall{10}} & {\recall{50}} & {Accuracy} \\
\midrule
\multicolumn{5}{l}{\itshape LFM2-2.6B} \\
\quad Backbone & {--} & {--} & {--} & 0.6461 \\
\quad No replay & \bfseries 0.2876 & \bfseries 0.3336 & \bfseries 0.2096 & 0.2489 \\
\quad Replay & 0.2871 & 0.3276 & 0.2027 & \bfseries 0.3185 \\
\addlinespace[3pt]
\multicolumn{5}{l}{\itshape Qwen3-14B-Base} \\
\quad Backbone & {--} & {--} & {--} & 0.8044 \\
\quad No replay & 0.2962 & \bfseries 0.3572 & 0.2207 & 0.2534 \\
\quad Replay & \bfseries 0.2969 & 0.3553 & \bfseries 0.2230 & \bfseries 0.3940 \\
\bottomrule
\end{tabular}

\end{table}

\subsubsection{Data mix}
\label{sec:pretraining-mix}
In addition to ablating the value of product attribute data and general-domain replay data to the final search and recommendation tasks (\cref{sec:pretraining-attribute-data,sec:pretraining-replay}),
we also evaluate whether the presence of recommendation data helps the search task performance and vice versa. We notice that recommendation data helps the search task, whereas search data helps
the recommendation task only slightly (\cref{tab:cross-task}).

Having gained an understanding of how the different data streams contribute, we adapt the baseline mix (\cref{tab:baseline-mix}) to the full-length training run: we cap how many times the smaller streams repeat and give the freed share of the loss tokens to the recommendation data and to the product attribute data of the full catalog. The final data mix is detailed in \cref{tab:mixture}.

\begin{table}[htbp]
  \caption{Cross-task transfer: LFM2-2.6B trained on the joint baseline mix
    (\cref{tab:baseline-mix}) and with either the search or recommendation
    streams removed. The other streams take the freed tokens in proportion.
    A dash marks the task whose training data were removed.
    Evaluation protocol E3 (\cref{tab:eval-protocols}).
    Best value per column in bold.}
  \label{tab:cross-task}
  \centering\footnotesize
\begin{tabular}{l S[table-format=1.4] S[table-format=1.4] S[table-format=1.4]}
\toprule
 & \multicolumn{1}{c}{Search relevance} & \multicolumn{1}{c}{Search purchases} & \multicolumn{1}{c}{Recommendation} \\
\cmidrule(lr){2-2} \cmidrule(lr){3-3} \cmidrule(lr){4-4}
Arm & {\onedcg{10}} & {\recall{10}} & {\recall{50}} \\
\midrule
Baseline & \bfseries 0.2876 & \bfseries 0.3336 & \bfseries 0.2096 \\
No search data & {--} & {--} & 0.2087 \\
No recommendation data & 0.2854 & 0.3263 & {--} \\
\bottomrule
\end{tabular}

\end{table}

\subsection{Training}
\label{sec:pretraining-training}

We start from three pre-trained backbone LLMs of different sizes: LFM2-2.6B \citep{liquid2025lfm2}, Qwen3-4B-Base and Qwen3-14B-Base \citep{yang2025qwen3}. To use the \sids as tokens, we
add one token per codeword of each \sid level (\cref{sec:tokenizer}) to the vocabulary of each backbone, as well as special tokens for the task formats (\cref{fig:training-examples}). We initialize
the rows added to the embedding matrices at the mean of the existing rows.

We train each model in two stages, which we describe in the following sections: embedding alignment and continued pre-training. In both stages, we use the AdamW optimizer
\citep{loshchilov2019adamw} and pack the training examples into fixed-length sequences. The settings of each stage are detailed in \cref{tab:training-recipe}.

\begin{table}[htbp]
  \caption{Training recipe per stage, with stage 2 split into the phases of its warmup-stable-decay schedule
    (\cref{sec:pretraining-cpt}). LR:
    learning rate. Warmup is linear. Replay: the general-domain replay data's share of the loss tokens
    (\cref{sec:pretraining-replay}). In both stages, AdamW has $\beta_1 = 0.9$ and $\beta_2 = 0.999$, gradients are
    clipped at norm 1.0, and a step holds 1,048,576 tokens in packed sequences of 4,096 tokens.}
  \label{tab:training-recipe}
  \centering\footnotesize
\begin{tabular}{l S[table-format=5] S[table-format=2.1] l S[table-format=4] l S[table-format=1.1] S[table-format=2]}
\toprule
Model & {Steps} & {Tokens (B)} & Peak LR & {Warmup steps} & Schedule & {Weight decay} & {Replay (\%)} \\
\midrule
\multicolumn{8}{l}{\itshape Stage 1: embedding alignment} \\
\quad All models & 4000 & 4.2 & \num{1e-3} & 200 & Cosine to zero & 0.0 & 0 \\
\addlinespace[3pt]
\multicolumn{8}{l}{\itshape Stage 2: continued pre-training, warmup and stable phases} \\
\quad LFM2-2.6B & 78000 & 81.8 & \num{5e-4} & 3000 & Constant & 0.1 & 0 \\
\quad Qwen3-4B-Base & 40000 & 41.9 & \num{1e-4} & 3000 & Constant & 0.1 & 10 \\
\quad Qwen3-14B-Base & 40000 & 41.9 & \num{1e-4} & 3000 & Constant & 0.1 & 10 \\
\addlinespace[3pt]
\multicolumn{8}{l}{\itshape Stage 2: continued pre-training, decay phase} \\
\quad LFM2-2.6B & 20000 & 21.0 & \num{5e-4} & 0 & Linear to zero & 0.1 & 0 \\
\quad Qwen3-4B-Base & 20000 & 21.0 & \num{1e-4} & 0 & Linear to zero & 0.1 & 0 \\
\quad Qwen3-14B-Base & 20000 & 21.0 & \num{1e-4} & 0 & Linear to zero & 0.1 & 0 \\
\bottomrule
\end{tabular}

\end{table}

\subsubsection{Embedding alignment}
\label{sec:pretraining-alignment}
The newly added \sid embeddings have not yet learned the meanings of their codewords. To give these new tokens meaningful embeddings before full training, we apply an embedding alignment step; training only the new embeddings while the backbone and its original embeddings stay frozen. This stage leverages only product attribute data (\cref{sec:pretraining-attribute-data}), in both directions: from an attribute to the \sid and from the \sid to an attribute. We train it without weight decay, so that the original embeddings stay exactly as they were.
We ablate the embedding alignment stage on LFM2-2.6B and notice that it improves both the search and recommendation metrics slightly (\cref{tab:embedding-alignment}), so we use it for all three models.


\begin{table}[htbp]
  \caption{Embedding alignment: both arms add the \sid tokens to the vocabulary of LFM2-2.6B and
    pre-train on the baseline mix (\cref{tab:baseline-mix}). With alignment, a first stage of 4,000 steps
    of 1,048,576 tokens trains only the new embeddings, on product documents; without it, all weights
    train from the start. Evaluation protocol E3 (\cref{tab:eval-protocols}). Better arm in bold.}
  \label{tab:embedding-alignment}
  \centering\footnotesize
\begin{tabular}{l S[table-format=1.4] S[table-format=1.4] S[table-format=1.4]}
\toprule
 & \multicolumn{1}{c}{Search relevance} & \multicolumn{1}{c}{Search purchases} & \multicolumn{1}{c}{Recommendation} \\
\cmidrule(lr){2-2} \cmidrule(lr){3-3} \cmidrule(lr){4-4}
Arm & {\onedcg{10}} & {\recall{10}} & {\recall{50}} \\
\midrule
Baseline & \bfseries 0.2876 & \bfseries 0.3336 & \bfseries 0.2096 \\
No embedding alignment & 0.2860 & 0.3268 & 0.2067 \\
\bottomrule
\end{tabular}

\end{table}

\subsubsection{Continued pre-training}
\label{sec:pretraining-cpt}
In the second stage, we train all weights on the final data mix (\cref{tab:mixture}). For the larger Qwen3 models, we also add the general-domain replay data (\cref{sec:pretraining-replay}). We choose the
peak learning rates with learning-rate sweeps on LFM2-2.6B and Qwen3-14B-Base. Qwen3-4B-Base diverged at higher learning rates, so we use the same peak learning rate as for Qwen3-14B-Base.

We use a warmup-stable-decay (WSD) learning-rate schedule \citep{hu2024minicpm,hagele2024scaling}: the learning rate warms up, stays constant, and finally decays linearly to zero. In our
learning-rate sweep on LFM2-2.6B, WSD matched or outperformed a cosine schedule at each peak learning rate up to the one we use. Since the constant phase has no fixed length, we can start the decay
from any of its checkpoints, resuming with the checkpoint's optimizer state. To choose the checkpoint, we evaluate a few candidates on the search and recommendation tasks and decay the best one. Given that we are
data-constrained on the search task, we observe that the search metrics tend to deteriorate late in the constant phase, so the best checkpoint to start decay from is not the latest one.

During the decay, we train on a fresh shuffle of the data and replace the recommendation streams with a single stream in the fine-tuning format (\cref{fig:training-examples}), keeping
recommendation's share of the tokens. For the Qwen3 models, we also drop the replay data. Metrics before and after the decay can be seen in \cref{tab:anneal}.

\begin{table}[htbp]
  \caption{Effect of the learning rate decay on search and recommendation metrics. Evaluation protocol E4 (\cref{tab:eval-protocols}). Best value per column in bold.}
  \label{tab:anneal}
  \centering\footnotesize
\begin{tabular}{l S[table-format=1.4] S[table-format=1.4] S[table-format=1.4]}
\toprule
 & \multicolumn{1}{c}{Search relevance} & \multicolumn{1}{c}{Search purchases} & \multicolumn{1}{c}{Recommendation} \\
\cmidrule(lr){2-2} \cmidrule(lr){3-3} \cmidrule(lr){4-4}
Arm & {\onedcg{10}} & {\recall{10}} & {\recall{50}} \\
\midrule
Constant learning rate, step 78,000 & 0.3108 & 0.3173 & 0.2075 \\
Decayed, step 98,000 & \bfseries 0.3211 & \bfseries 0.3415 & 0.2142 \\
Decayed, fine-tuning-format recommendation data & 0.3194 & 0.3392 & \bfseries 0.2533 \\
\bottomrule
\end{tabular}

\end{table}


\FloatBarrier

\section{Post-training}
\label{sec:post-training}

Post-training has two goals: improve judged search quality and combine task-specific capabilities in
one model. We first study adapting the multi-token prediction generator using feedback from a relevance
judge. We then investigate task unification starting from the autoregressive model
(\cref{sec:post-training-unification}). These are separate experiments; we evaluate the search-quality
methods on search only.

\subsection{Improving search quality}
\label{sec:post-training-search}
Pre-training builds catalog knowledge and learns patterns in search and buyer interactions. For a given
query, however, many products may be relevant, including products absent from the recorded interactions.
Post-training lets us give the generator more direct supervision about which products satisfy the query
and the buyer's preferences. We compare three ways to use this supervision: learning from positively
judged candidates, learning preferences between candidates, and obtaining feedback on candidates
generated during training.

All three methods start independently from the same LFM2-2.6B checkpoint after continued pre-training
and learning-rate decay (CPT+MTP). A fine-tuned Qwen3.5-2B judge scores products against the query and
buyer profile, when available. For the offline methods, candidates come from search engine results
pages (SERPs) and, in some variants, the CPT+MTP generator. We deduplicate candidates by \sid and retain
the highest judge score. Both offline objectives use a relevance token in the prompt to distinguish
the desired output from purchase prediction. Data construction, loss definitions and training settings
are detailed in \cref{sec:appendix-search-post-training}.

\subsubsection{Supervised fine-tuning}
\label{sec:sft}
Soft-target cross-entropy (CE) trains the model to generate \sids of high-scoring products. We retain
candidates meeting a score threshold and weight them by their judge scores, so several products can
provide targets for the same query. At each \sid position, a codeword's target probability is the sum
of the weights of products containing it. The MTP model learns these distributions in parallel.

\subsubsection{Preference optimisation}
\label{sec:preference}
\paragraph{Offline CE+DPO.}
Direct preference optimisation (DPO; \citealt{rafailov2023dpo}) adds supervision from low-scoring
candidates as well as positive ones. We combine positive-target CE with a preference loss that favors
higher-scoring products over lower-scoring ones, relative to the frozen CPT+MTP reference. The
candidate weights depend on distance from the positive--negative threshold. Queries with positive
candidates but no negatives still contribute CE. We compare SERP-only
training with a variant that also includes generated candidates.

\paragraph{GRPO.}
Group relative policy optimisation (GRPO; \citealt{shao2024deepseekmath}) obtains judge feedback on
candidates generated as training proceeds. For each query, constrained MTP beam traversal proposes
32 \sids. We score the first product returned when resolving each \sid and use the judge's score as
its reward. Rewards are
centered within the query group, and a clipped policy objective with a KL penalty against the initial
model updates the generator.

Our implementation uses ranked proposals rather than independent policy samples, a modified reward
normalization, and asynchronous generation and training. We do not apply importance correction for
the lag between the generation and training policies. Here, feedback during training comes from the
judge; it does not require live buyer interactions. The complete procedure is given in
\cref{sec:appendix-post-training-grpo}.

\subsubsection{Results and interpretation}
\label{sec:post-training-results}
\Cref{tab:post-training} compares the generators under protocol E5, without a ranker. All three
approaches improve \onedcg{10} over CPT+MTP, with GRPO achieving the highest observed score. Adding
DPO on SERP candidates provides only a small improvement over CE. Including generated candidates
increases the score further, but also changes the data and training schedule, so that comparison
does not isolate the effect of the preference loss. Increasing the DPO coefficient adds no
practically meaningful improvement.

\begin{table}[htbp]
  \caption{Search post-training from the same CPT+MTP initialization, without a ranker. Evaluation
    protocol E5 (\cref{tab:eval-protocols}). All-source CE+DPO includes generated candidates. Data and
    training budgets differ across methods. A dash indicates purchase recall not reported for that
    arm. Best value per column in bold.}
  \label{tab:post-training}
  \centering\small
\begin{tabular}{l S[table-format=1.4] S[table-format=1.4]}
\toprule
 & \multicolumn{1}{c}{Search relevance} & \multicolumn{1}{c}{Search purchases} \\
\cmidrule(lr){2-2} \cmidrule(lr){3-3}
Recipe & {\onedcg{10}} & {\recall{10}} \\
\midrule
CPT+MTP & 0.2811 & \bfseries 0.2840 \\
Soft-target CE, SERP & 0.2930 & 0.2767 \\
CE+DPO, SERP & 0.2937 & {--} \\
CE+DPO, all sources & 0.2950 & 0.2745 \\
CE+DPO, all sources, DPO weight 2 & 0.2950 & {--} \\
GRPO & \bfseries 0.2970 & 0.2782 \\
\bottomrule
\end{tabular}

\end{table}

Purchase recall declines despite the improvement in \onedcg{10}. These metrics capture different
aspects of retrieval. Purchase recall measures recovery of observed purchases and is affected by
exposure bias: relevant products that buyers were never shown cannot appear as positive targets.
Improving judged quality can therefore involve retrieving alternatives that receive no credit under
purchase recall. We interpret the two metrics together; the observed decline does not establish
whether these alternatives would lead to more or fewer purchases if shown.
\FloatBarrier

\subsection{Unifying search and recommendation}
\label{sec:post-training-unification}
Search and recommendation depend on shared knowledge of products and buyer preferences. A unified model
would let us train and update these capabilities together, while serving requests with different kinds of
context. The challenge is to preserve the quality available from task-specific post-training. We study
this question starting from the autoregressive CPT (\cref{tab:ar-mtp}), which has already been pre-trained
on both tasks.

\subsubsection{The cost of joint fine-tuning}
\label{sec:unification-setup}
Our baseline is joint supervised fine-tuning (joint SFT): 5,000 steps on a mix of recommendation and
search examples. We compare it with experts fine-tuned on each task separately
(\cref{tab:distillation-hparams}). At the same step budget, the recommendation expert improves
\recall{50} by 1.7\% over joint SFT, but loses search quality. The search expert has no detectable
advantage over joint SFT on search and substantially reduces recommendation quality
(\cref{tab:unification}). We call the recommendation gap the \emph{unification tax}: mixing the
fine-tuning data leaves some of the specialist's gains unrealized.

Our goal is therefore to recover the recommendation gains while retaining joint SFT's search quality.
Weight merging \citep{ilharco2023taskarithmetic} does not resolve this trade-off in our experiments
(\cref{sec:appendix-merging,fig:merge-frontier}), motivating an approach that learns from each expert
on its own task.

\subsubsection{Distilling task experts into one model}
\label{sec:unification-distillation}
\begin{figure}[t]
  \centering
  \IfFileExists{figures/pipeline_overview.pdf}{\includegraphics[width=0.8\linewidth]{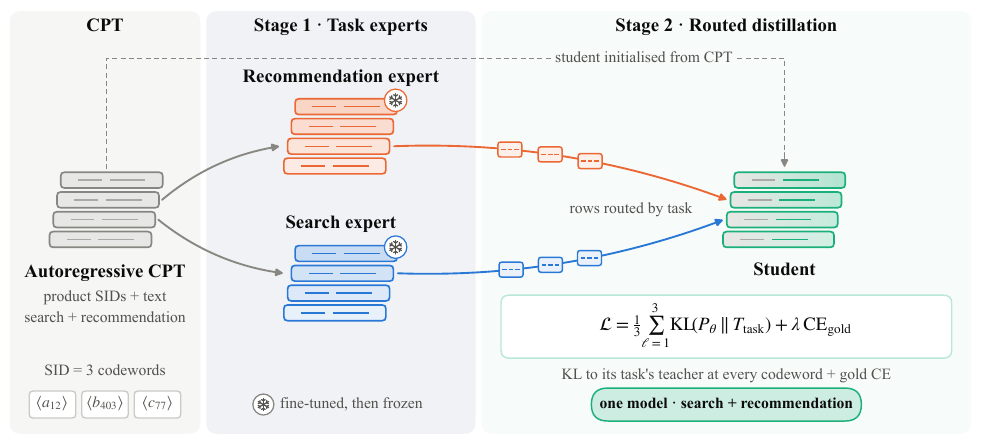}}{\todo{figure pending}}
  \caption{Distilling task experts into a shared model. Each expert is fine-tuned from the CPT and
    frozen. A student initialized from the same CPT learns from examples of both tasks, with each
    example routed to its task's expert.}
  \label{fig:post-training-pipeline}
\end{figure}

We train and freeze an expert for each task, then distill both into a student initialized from the CPT
(\cref{fig:post-training-pipeline}). Each training example is routed to its task's expert. A search
example receives guidance from the search expert, and a recommendation example from the recommendation
expert. The experts' distributions provide supervision over alternative codewords as well as the labeled
product, giving the shared student a way to learn their task-specific predictions
\citep{hinton2015distillation,ma2026mopd}.

The loss combines reverse KL divergence to the expert with cross-entropy on the labeled product's
\sid. The first term transfers the expert's predictions; the second anchors training to the labeled
target. For a prompt $x$, task $k$, gold \sid $y=(y_1,y_2,y_3)$ and frozen expert $T_k$, the student
$P_\theta$ minimizes
\begin{equation}
  \label{eq:distillation}
  \mathcal{L}(\theta)=\frac{1}{3}\sum_{\ell=1}^{3}
    \mathrm{KL}\bigl(P_\theta(\cdot\mid x,y_{<\ell})\,\big\|\,T_k(\cdot\mid x,y_{<\ell})\bigr)
    -\frac{\lambda}{3}\sum_{\ell=1}^{3}\log P_\theta(y_\ell\mid x,y_{<\ell}),
  \qquad \lambda=0.25.
\end{equation}
Both models condition on the gold prefix, and the KL is exact over the full vocabulary. Teacher
selection, training settings and the objective and prefix comparisons are described in
\cref{sec:appendix-teachers,sec:appendix-two-stage,sec:appendix-unification-details}.

\subsubsection{Results and implications}
\label{sec:unification-results}
\label{sec:unification-eval}
\begin{figure}[t]
  \centering
  \IfFileExists{figures/unification_frontier.pdf}{\includegraphics[width=\linewidth]{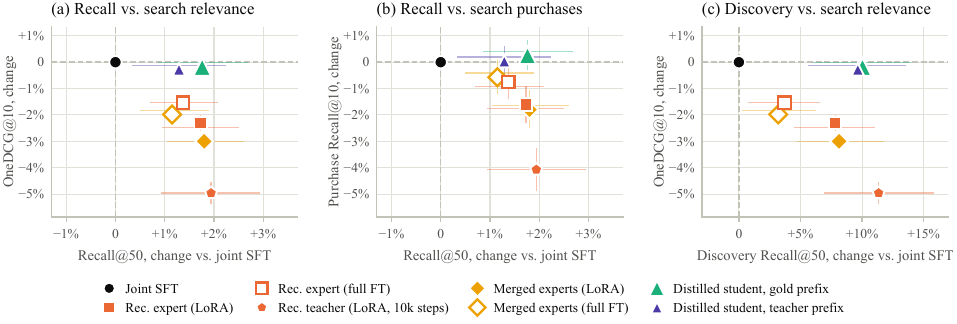}}{\todo{figure pending}}
  \caption{Search against recommendation Pareto frontier, as change (\%) from joint SFT at the origin, with 95\%
    paired-bootstrap whiskers: (a) \recall{50} against \onedcg{10}, (b) \recall{50} against purchase
    \recall{10}, (c) discovery \recall{50} against \onedcg{10}. Hollow markers: full fine-tuning. For a full sweep over model merges see \cref{fig:merge-frontier}.}
  \label{fig:unification-frontier}
\end{figure}

\Cref{tab:unification} reports absolute scores for recommendation \recall{50} and discovery
\recall{50}, and for search purchase \recall{10} and \onedcg{10} under protocol E5.
\Cref{fig:unification-frontier} shows relative changes from joint SFT with 95\% paired-bootstrap
intervals; a change is significant when its interval excludes zero. Cohorts and evaluation details
are given in \cref{sec:appendix-unification-eval}.

\begin{table}[t]
  \caption{Absolute metric performance across model variations, the first row (CPT) is the  base checkpoint all students and experts start from, second row is the effect of training with a 1:1:1 dataset mixture. Merged experts shown use task arithmetic, weights (1,1).
}
  \label{tab:unification}
  \centering\footnotesize\setlength{\tabcolsep}{2pt}
  \IfFileExists{tables/unification.tex}{
\begin{tabular}{l c c c c}
\toprule
 & \multicolumn{2}{c}{Recommendation} & \multicolumn{2}{c}{Search} \\
\cmidrule(lr){2-3} \cmidrule(lr){4-5}
Model & {\recall{50}} & {\shortstack{Discovery\\\recall{50}}} & {\shortstack{Purchase\\\recall{10}}} & {\shortstack{Relevance\\\onedcg{10}}} \\
\midrule
CPT & \num{0.1939} & \num{0.0288} & \num{0.3246} & \bfseries\num{0.3083}\\
\midrule
Joint SFT & \num{0.2411} & \num{0.0423} & \num{0.3259} & \num{0.3061} \\
\midrule
Rec. expert (LoRA, 5k steps) & \num{0.2453} & \num{0.0456} & \num{0.3206} & \num{0.2990} \\
Rec. expert (LoRA, 10k steps) & \bfseries\num{0.2458} & \bfseries\num{0.0471} & \num{0.3126} & \num{0.2909} \\
Search expert (full FT) & \num{0.1924} & \num{0.0287} & \num{0.3257} & \num{0.3059} \\
\midrule
Merged experts (LoRA) & \num{0.2454} & \num{0.0457} & \num{0.3200} & \num{0.2969} \\
Merged experts (full FT) & \num{0.2439} & \num{0.0437} & \num{0.3240} & \num{0.3000} \\
\midrule
Distillation, gold prefix & \num{0.2453} & \num{0.0466} & \bfseries\num{0.3267} & \num{0.3055} \\
Distillation, teacher prefix & \num{0.2442} & \num{0.0464} & \num{0.3260} & \num{0.3053} \\
\bottomrule
\end{tabular}

}{\todo{table pending}}
\end{table}

The distilled student improves \recall{50} by 1.76\% and discovery \recall{50} by 10.03\% over joint
SFT, with no detected change on either search metric (\cref{tab:unification,fig:unification-frontier}).
It recovers much of the recommendation teacher's gain while avoiding the search losses observed when
using that teacher directly. Discovery is important here because overall recall is dominated by
purchases of products already in the buyer's history. A breakdown on the same users attributes most
of the student's overall recall gain to purchases of products new to the buyer
(\cref{tab:discovery-retargeting}).

These results support task-routed distillation as a route to a shared model that retains the benefits
of specialization. They do not isolate distillation from the additional training budget and teacher
choices, and the intervals still allow small search losses. Each model was trained once on a development
benchmark that also guided teacher and recipe selection, so the intervals are nominal. Distilling
directly from the CPT removes the joint-SFT stage, although an earlier two-stage recipe performs better
on search purchase recall (\cref{sec:appendix-two-stage}). Neither recipe recovers the CPT's full
\onedcg{10}, leaving room to improve the search teacher or apply search-specific post-training
(\cref{sec:preference}). The unified student remains experimental; production uses the task-specific
models described in \cref{sec:integration}.
\FloatBarrier

\section{Integration}
\label{sec:integration}

\subsection{Search}
\label{sec:integration-search}
\sidllm plays two roles in the Shopify search stack: it is one of the retrievers, and it supplies features to the
rankers. Keyword retrieval, embedding retrieval and \sidllm each contribute candidates, and two learned ranking
stages, L2 and L3, then order the merged set. \sidllm contributes its top 100 \sids for the query and the buyer's
profile (\cref{sec:inference}), each resolved to one product (\cref{sec:tokenizer}).

\sidllm scores only the products it generates, so we turn its output into features that every candidate has,
whichever retriever found it. For each candidate, we compute its cosine similarity to each of \sidllm's top 50
products, with embeddings from the in-house model the \sids are built from (\cref{sec:tokenizer-data}). The
maximum, mean and minimum of these similarities are features of both rankers.

\subsection{Recommendation}
\label{sec:integration-rec}
In Shopify's recommendation stack, \sidllm is one candidate generator among several. For each buyer, it generates 200
candidate products from their interaction history, using the same prompt format it was fine-tuned on. These
candidates are pooled with those from the other generators, and the downstream ranking stages order the combined
pool. The production model is the AR LFM2-2.6B, paired with the joint-probability decoder described below.

\paragraph{From semantic IDs to products.} A \sid names a cluster of similar products rather than a single
product: of the \sids the model predicts, 78.9\% are shared by more than one product. The model can therefore
predict the right \sid and still miss the purchase, if the decode picks the wrong product from the cluster. Instead
of taking one product per \sid (\cref{sec:tokenizer}), we let the most confident \sids contribute more: we score each
candidate product by its joint probability with the user. The beam gives $\log p(s \mid u)$, the summed
log-probability of the \sid's tokens. Within the \sid's cluster, we take
$p(i \mid s) = \mathrm{pop}_i / \sum_{j \in s} \mathrm{pop}_j$, where $\mathrm{pop}$ is a popularity score. A product
scores $\log p(s \mid u) + \log p(i \mid s)$, and the decode keeps the 200 highest-scoring products, considering up to
ten per \sid. A popular product from a confident \sid can therefore outrank the top product of a \sid further down the
beam. The score has no tuning parameter.

We compare decodes with the model and its 200 beam \sids held fixed (\cref{tab:rec-decode}). Joint scoring improves
\recall{50}, both over one product per \sid and over a fixed schedule that gives the three highest-ranked \sids two
products each. Against one product per \sid, it raises \recall{50} by 15.8\%.
\begin{table}[htbp]
  \caption{Relative change of each \sid-to-product decode against one product per \sid, for the production
    \sidllm with its 200 beam \sids held fixed, evaluated on 9,990 held-out purchasers. Best value in bold.}
  \label{tab:rec-decode}
  \centering\small
\begin{tabular}{l S[table-format=2.1]}
\toprule
 & \multicolumn{1}{c}{Recall change (rel. \%)} \\
\cmidrule(lr){2-2}
Decode & {\recall{50}} \\
\midrule
One product per SID & {--} \\
Two products for the top 3 SIDs & 6.7 \\
Joint probability & \bfseries 15.8 \\
\bottomrule
\end{tabular}

\end{table}

\section{Inference and Deployment}
\label{sec:inference}

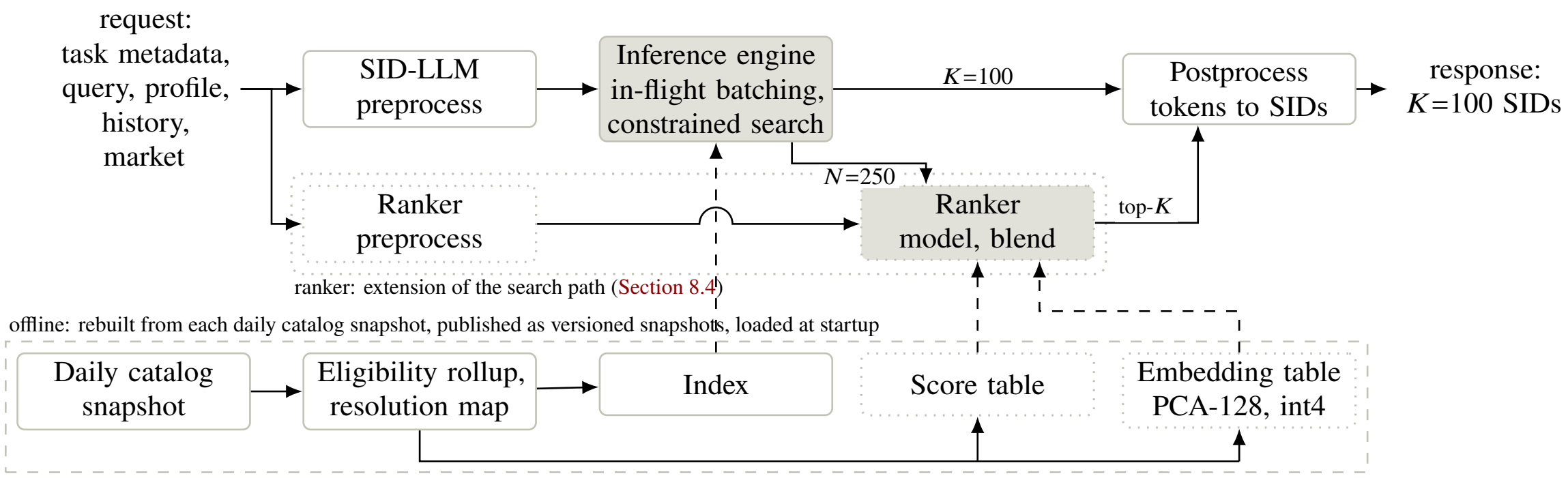
\begin{figure}[htb]
  \centering
  \begin{tikzpicture}[
    font=\scriptsize, node distance=6mm,
    stage/.style={draw=axis, line width=0.5pt, rounded corners=2pt, align=center, inner sep=2.2pt,
      minimum height=6mm, text width=21mm},
    engine/.style={stage, fill=grid},
    ext/.style={dotted, line width=0.6pt},
    endpoint/.style={align=center, text width=16mm},
    group/.style={draw=axis, dotted, line width=0.6pt, rounded corners=3pt, inner sep=3pt},
    band/.style={draw=axis, dashed, line width=0.5pt, inner sep=3pt},
    note/.style={font=\tiny, inner sep=1pt},
    tag/.style={note, fill=white},
    edge/.style={-{Latex[length=1.6mm, width=1.3mm]}, line width=0.5pt},
    load/.style={edge, dashed}]
    \node[endpoint] (req) {request:\\task metadata,\\query, profile,\\history, market};
    \node[stage, right=of req] (pre) {\sidllm\\preprocess};
    \node[engine, right=of pre] (eng) {Inference engine\\in-flight batching,\\constrained search};
    \node[stage, right=28mm of eng] (post) {Postprocess\\tokens to \sids};
    \node[endpoint, right=3mm of post] (resp) {response:\\$K{=}100$ \sids};
    \draw[edge] (req) -- (pre);
    \draw[edge] (pre) -- (eng);
    \draw[edge] (eng) -- node[tag, above] {$K{=}100$} (post);
    \draw[edge] (post) -- (resp);
    \node[stage, ext, anchor=north] (rpre) at ($(pre.south)+(0,-5.5mm)$) {Ranker\\preprocess};
    \coordinate (gap) at ($(eng.east)!0.5!(post.west)$);
    \node[engine, ext, anchor=north] (rank) at (gap |- rpre.north) {Ranker\\model, blend};
    \node[group, fit=(rpre)(rank)] (extension) {};
    \node[tag, anchor=north west] at (extension.south west) {ranker: extension of the search path (\cref{sec:inference-ranker})};
    \draw[edge] (req.east) -- ++(3mm,0) |- (rpre);
    \draw[edge] let \p1 = (eng.south), \p2 = (rpre.east) in
      (\p2) -- (\x1 - 1.6mm, \y2) arc[start angle=180, end angle=0, radius=1.6mm] -- (rank.west |- \p2);
    \coordinate (nout) at ([xshift=-4mm]eng.south east);
    \coordinate (nin) at ([xshift=-5mm]rank.north);
    \coordinate (jog) at ($(eng.south)!0.5!(rank.north)$);
    \draw[edge] (nout) -- (nout |- jog) -- node[tag, below] {$N{=}250$} (nin |- jog) -- (nin);
    \draw[edge] (rank.east) -| node[tag, pos=0.25, above] {top-$K$} ([xshift=-4mm]post.south);
    \coordinate (row) at ($(extension.south)+(0,-7.5mm)$);
    \node[stage, anchor=north] (idx) at (row -| eng) {Index};
    \node[stage, anchor=north east] (elig) at ($(idx.north west)+(-6mm,0)$) {Eligibility rollup,\\resolution map};
    \node[stage, anchor=north east] (snap) at ($(elig.north west)+(-5mm,0)$) {Daily catalog\\snapshot};
    \node[stage, ext, anchor=north] (aux) at (row -| rank) {Score table};
    \node[stage, ext, anchor=north] (emb) at (row -| post) {Embedding table\\PCA-128, int4};
    \coordinate (bus) at ($(elig.south)+(0,-3mm)$);
    \draw[edge] (snap) -- (elig);
    \draw[edge] (elig) -- (idx);
    \draw[line width=0.5pt] (elig.south) -- (bus) -- (bus -| emb.south);
    \draw[edge] (bus -| aux.south) -- (aux.south);
    \draw[edge] (bus -| emb.south) -- (emb.south);
    \node[band, fit=(snap)(elig)(idx)(aux)(emb)(bus)] (offline) {};
    \node[note, anchor=south west] at (offline.north west) {offline: rebuilt from each daily catalog snapshot, published as versioned snapshots, loaded at startup};
    \draw[load] (idx) -- (eng);
    \draw[load] (aux.north) -- (aux.north |- rank.south);
    \draw[load] (emb.north) -- ++(0,4mm) -| ([xshift=6mm]rank.south);
  \end{tikzpicture}
  \caption{Search serving path (top) and artifact pipeline (bottom). Shaded: engine processes;
    dotted: the ranker extension of the search path (\cref{sec:inference-ranker});
    dashed: artifacts loaded at startup.}
  \label{fig:serving-path}
\end{figure}

\sidllm generates candidates for production search and recommendation stacks
(\cref{sec:integration}). This section describes the online search serving path, which uses MTP
with a learned ranker. Recommendation uses an AR model and joint-probability product resolution
(\cref{sec:integration-rec}). The catalog holds billions of products; hundreds of millions of
them are eligible at any time and resolve to about $4\times10^{8}$ distinct \sids
(\cref{sec:tokenizer}), and a sixth of product records change daily. Serving has strict
latency requirements: each search request must return its valid \sids within tens of milliseconds
on a fraction of one GPU (\cref{fig:serving-path}). We build on STATIC \citep{su2026static},
which flattens the catalog trie into a compressed-sparse-row index on the accelerator, so that
constraining a decoding step is a vectorized gather rather than a host round trip; in our
setting, this reduces the constraint overhead from 18--33 to about 1\,ms per request. We
extend the index from one catalog-wide constraint to per-request eligibility.

\subsection{Inference engine}
\label{sec:inference-engine}

Generative retrieval pairs variable-length inputs with wide, shallow, constrained decoding
rather than long sequential generation. Complete serialized prompts range from six tokens to
an enforced limit of 1,500 in \sidllm's tokenizer and vary widely between tasks: a
non-personalized search prompt is a short query, a personalized one adds a user profile, and
a recommendation prompt encodes up to 100 interaction events (\cref{fig:training-examples}).
For search, decoding is a constrained beam search of width $W{\le}250$, followed by re-ranking
to return $K{=}100$ candidates, each represented by three output codewords. We enforce the decoding constraints
through a CUDA/C++ plugin inside the the inference engine.

\begin{table}[t]
  \caption{End-to-end server-side latency (ms) on one B200 3g.90gb partition under simulated
    production search traffic: the same 37,924 requests per rate and arm; all 455,088 succeeded.
    The third arm adds the ranker (\cref{sec:inference-ranker}).}
  \label{tab:inference-latency}
  \centering\small
\begin{tabular}{S[table-format=3] S[table-format=2.1] S[table-format=2.1] S[table-format=3.1] S[table-format=1.1] S[table-format=2.1] S[table-format=2.1] S[table-format=2.1] S[table-format=2.1] S[table-format=3.1]}
\toprule
 & \multicolumn{3}{c}{AR, beam 100} & \multicolumn{3}{c}{MTP, beam 250} & \multicolumn{3}{c}{MTP + ranker, 250 to 100} \\
\cmidrule(lr){2-4} \cmidrule(lr){5-7} \cmidrule(lr){8-10}
{Offered requests/s} & {p50} & {p95} & {p99} & {p50} & {p95} & {p99} & {p50} & {p95} & {p99} \\
\midrule
25 & 19.2 & 31.2 & 38.9 & 9.7 & 18.2 & 22.2 & 17.5 & 28.2 & 34.0 \\
50 & 21.2 & 36.5 & 50.5 & 9.4 & 17.6 & 22.8 & 17.5 & 29.2 & 38.2 \\
75 & 24.0 & 45.3 & 161.0 & 9.2 & 17.8 & 25.5 & 17.8 & 31.8 & 59.3 \\
100 & 28.2 & 63.3 & 568.7 & 9.1 & 18.4 & 61.2 & 18.5 & 36.6 & 231.4 \\
\bottomrule
\end{tabular}

\end{table}

\subsection{Decoding strategy}
\label{sec:inference-decoding}

Autoregressive (AR) decoding requires one backbone pass per codeword (the prefill, then two
incremental passes), each conditioned on the codewords chosen so far. Multi-token prediction
(MTP, \cref{sec:sid-llm-mtp}) predicts all three codeword distributions in one prefill, each
conditioned on the request prefix rather than on earlier codewords; the index constrains their
combinations, and a beam walk over it assembles candidates.
\Cref{tab:inference-latency} reports end-to-end server-side latency on simulated production
search traffic across offered loads.

In offline evaluation, MTP scores below AR on both top-10 metrics, and widening its beam from
100 to 250 does not close the gap (\cref{tab:ar-mtp}). For search, the ranker
(\cref{sec:inference-ranker}) more than makes up the difference, raising both metrics above AR.

\subsection{Catalog index}
\label{sec:inference-artifacts}

Each day we rebuild the index from a fresh catalog snapshot
(\cref{fig:serving-path}, bottom). Because several products can share one \sid, an
eligibility rollup makes a \sid eligible in every market where at least one of its products
is. In the snapshot described here, 354M \sids are eligible in at least one market and can be
retrieved. Each serving process loads one validated snapshot at startup, so eligibility is current
as of that snapshot until redeployment. Because a request's market determines which
\sids are valid, the index also carries compact membership representations that
preserve shared prefixes, so we can express which products satisfy each constraint without
duplicating the underlying catalog structure; we describe it only functionally.

\subsection{Ranker}
\label{sec:inference-ranker}

The ranker extends only the search path; recommendation resolves AR-generated \sids into 200
products using joint-probability scoring (\cref{sec:integration-rec}). For search, the generator returns
$N{=}250$ candidates, and a 350M-parameter LFM2
model, served as a second engine process, re-ranks them jointly in one prefill in which the query
and profile attend causally and the candidate embeddings bidirectionally. The ranker's two artifacts,
a per-\sid static quality score (the auxiliary score $a$) and a 128-dimensional int4 product-embedding
table keyed by \sid, come from the same daily pipeline (\cref{fig:serving-path}, bottom); both models
share one 89.0\,GiB GPU partition (\cref{fig:inference-memory} in the appendix).

We distill a relevance judge model with a Qwen3.5-9B backbone into the ranker. For each training query, the judge reads the query, the buyer profile when there is one and the
product text, and gives $y_i\in[0,1]$, its relevance
score for candidate $i$ of the 250 from \cref{tab:ar-mtp}'s MTP generator; the ranker sees each product
only as its embedding $\bm{e}_i$. The ranker's score $r_i$ combines two signals: a relevance estimate
from the candidate's final hidden state, which has attended to the query, the profile and the other
candidates, and an embedding match between $\bm{e}_i$ and a query vector $\hat{\bm{q}}$ that the ranker
predicts from the query and profile. Training minimizes $\mathcal{L}=\mathcal{L}_{\mathrm{BCE}}+\mathcal{L}_{\mathrm{LL}}+\mathcal{L}_{\cos}$. The binary
cross-entropy $\mathcal{L}_{\mathrm{BCE}}$ fits $r_i$ to $y_i$ on the candidates the judge could score. The
LambdaLoss term $\mathcal{L}_{\mathrm{LL}}$ \citep{wang2018lambdaloss} trains the order: it penalizes pairs that $3r_i+a_i$ orders differently from $3y_i+a_i$, the mix of relevance and
static score in \cref{sec:eval-ndcg}'s gain, weighted by their effect on NDCG over the 100 candidates
returned. The cosine term $\mathcal{L}_{\cos}$ makes that embedding match servable. The product-embedding
model's query--product similarity alone already ranks candidates well, and an ablation without it
degrades the ranker markedly, but only the product embeddings are cached; embedding each query would
keep that model in memory and add its latency. $\mathcal{L}_{\cos}$ therefore pulls
$\hat{\bm{q}}$ toward that model's own query embedding $\bm{q}$, projected like the table into 128
dimensions, so $\hat{\bm{q}}^\top\bm{e}_i$ approximates the similarity without running the query encoder. We keep the checkpoint with the
lowest held-out loss over 10 epochs.

The served order sorts candidates by $0.5\tilde r+0.3\tilde a+0.2\tilde s_{\mathrm{MTP}}$, where $r$
is the ranker's score, $s_{\mathrm{MTP}}$ the generator's log-probability of
the candidate, and the tilde denotes min--max normalization over the request's candidates. Because
the auxiliary score is computed like the relevance metric's static score (\cref{sec:eval-ndcg}),
purchase recall is the more independent of the two metrics; both measure the whole blend. MTP with
the ranker scores above AR on both metrics (\cref{tab:ar-mtp}) and is faster than AR at every
percentile and offered rate, increasingly so under load (\cref{tab:inference-latency}); we therefore
serve MTP with the ranker.

\FloatBarrier

\section{Evaluation Results}
\label{sec:evaluation}

\subsection{Search}
\label{sec:eval-search}

\subsubsection{Online}
\label{sec:eval-search-online}
We evaluated several successive versions of \sidllm through a series of A/B tests in Shopify's search stack. The
treatment arms included both its candidate retrieval and the features it supplies to downstream rankers
(\cref{sec:integration-search}). We assessed success using purchases, high-quality clicks and clicks, measured both
in aggregate and separately for head, torso and tail queries, with these segments defined by query frequency.

The results of these evaluations supported deploying \sidllm in production as part of Shopify's search stack.

\subsubsection{Offline}
\label{sec:eval-search-offline}
To measure what \sidllm adds to search, we compare three configurations of Shopify's search stack
(\cref{sec:integration-search}): the full stack, the stack without \sidllm, and \sidllm alone.
Removing \sidllm disables both its candidate retrieval and the features it supplies to downstream
rankers. The \sidllm-only configuration retains its own ranker (\cref{sec:inference-ranker}),
but bypasses the other retrievers and the downstream L2 and L3 rankers.

With its own built-in ranker, \sidllm alone achieves search evaluation scores close to that of the entire
multi-model search stack without it (\cref{tab:search-stack}). The full stack performs best,
showing that \sidllm also adds value alongside the existing retrieval and ranking components.

\begin{table}[htbp]
    \caption{Search-stack ablation on the combined evaluation and on personalized search only.
  Changes are relative to the full stack with \sidllm within each setting.
  Evaluation protocol E6 (\cref{tab:eval-protocols}).}

  \label{tab:search-stack}
  \centering\small
\begin{tabular}{l S[table-format=1.5] S[table-format=-2.1] S[table-format=1.5] S[table-format=-1.1]}
\toprule
 & \multicolumn{2}{c}{Combined} & \multicolumn{2}{c}{Personalized only} \\
\cmidrule(lr){2-3} \cmidrule(lr){4-5}
Configuration & {\onedcg{10}} & {Change (\%)} & {\onedcg{10}} & {Change (\%)} \\
\midrule
Full stack with \sidllm & \bfseries 0.32690 & {--} & \bfseries 0.35718 & {--} \\
Full stack without \sidllm & 0.30570 & -6.5 & 0.32979 & -7.7 \\
\sidllm only & 0.28500 & -12.8 & 0.33608 & -5.9 \\
\bottomrule
\end{tabular}

\end{table}

\subsection{Recommendation}
\label{sec:eval-rec}

\subsubsection{Offline}
\label{sec:eval-rec-offline}

\paragraph{Pool recall.}
\label{sec:eval-rec-pool}
\sidllm is one candidate generator among several (\cref{sec:integration-rec}). What matters is what it adds to the
pool of candidates the ranking stages see. Each candidate generator contributes its top 200 products per user, and
the pool is the union of all generators' candidates and can contain more than 200 products. Pool recall is the share
of a user's orders in a held-out 7-day window that appear in the pool. It is averaged over users with at least one
such order.

\paragraph{Incremental pool recall.} To measure one generator's contribution, we remove it from the pool and record
how much pool recall drops. This leave-one-out drop is the generator's incremental pool recall. It counts the
purchases that only that generator finds.

\paragraph{Results.}
We put our best \sidllm model, the LFM2-2.6B with the joint-probability decode (\cref{sec:integration-rec}), into the
production pool next to the other production candidate generators (\cref{tab:rec-pool}). We evaluate on 70,747 users
with a purchase in the held-out week.
We compare it with recent engagement, which recommends products the buyer has recently interacted with, and HSTU
\citep{zhai2024hstu}, a generative recommender over item IDs.

\sidllm adds more to the pool than any other generator. Its incremental pool recall is 2.65 points, 61\% more than
recent engagement, the second-best candidate generator by incremental pool recall (1.64 points). It is also the best
at retrieval, with the highest \recall{50} of all production candidate generators.
\begin{table}[htbp]
  \caption{\sidllm in the production candidate pool. All values are in \%, and
    incremental pool recall is in percentage points of pool recall, with 200 candidates per generator.}
  \label{tab:rec-pool}
  \centering\small
\begin{tabular}{l S[table-format=2.2] S[table-format=1.2]}
\toprule
Candidate generator & {\recall{50}} & {Incremental pool recall} \\
\midrule
\sidllm (LFM2-2.6B) & \bfseries 24.61 & \bfseries 2.65 \\
Recent engagement & 17.28 & 1.64 \\
HSTU & 17.21 & 0.78 \\
\bottomrule
\end{tabular}

\end{table}

\subsubsection{Online}
\label{sec:eval-rec-online}
We ran an A/B test on our home feed using \sidllm with the joint-probability decoder
(\cref{sec:integration-rec,tab:rec-ab}). On the impacted surface, adding \sidllm raised
attributed orders by 7.95\%, converting users by 8.60\% and conversion rate by 9.85\%. All three lifts are significant after Dunnett correction \citep{dunnett1955}.
\begin{table}[htbp]
  \caption{A/B test in the home feed, the \sidllm arm against control. Every lift is significant after Dunnett correction.}
  \label{tab:rec-ab}
  \centering\small
\begin{tabular}{l S[table-format=1.2] S[table-format=1.2] S[table-format=1.2]}
\toprule
 & \multicolumn{3}{c}{Lift (\%)} \\
\cmidrule(lr){2-4}
Surface & {Orders} & {Converters} & {CVR} \\
\midrule
Impacted surface & 7.95 & 8.60 & 9.85 \\
\bottomrule
\end{tabular}

\end{table}

\section{Conclusion and Future Directions}
\label{sec:conclusion}
In this report, we presented \sidllm, Shopify's generative search and recommendation model.  We study the training stages and model variants used for search and recommendation, including \sid tokenization,
pre-training data and methodology, and post-training. Production search uses multi-token prediction combined with
a lightweight ranker to improve retrieval quality while keeping latency low. Production recommendation uses an
autoregressive model with joint-probability product resolution. Separately, we investigate task unification through
routed multi-teacher distillation with gold prefixes, producing an experimental unified student. Adding \sidllm
provides significant gains in Shopify's end-to-end search and recommendation systems.

Our goal for the future is to incorporate more of our discovery stack into the one generative model. The results comparing the \sidllm model to an entire
traditional search stack are promising, and in the future we anticipate product search to move more and more towards a single model. Additionally, we are training the model to do more tasks, such as personalized related-product recommendation and personalization taking into account in-session context.

Future work includes test-time scaling using Chain-of-Thought reasoning, and especially making it efficient enough to deploy in production. Additionally,
\sidllm currently can only ingest text and \sids, and we anticipate multi-modal understanding of products to provide further gains and avenues of research.

\section*{Acknowledgements}
The authors would like to thank Aaron Yuen and Sarthak Shah for driving product direction and coordinating the effort. Thanks to Linjun Yang for leadership in the project.  Thanks to Antoine Atallah for early support on the project. Thanks to Edoardo Mosca for guidance and management on the inference engine effort. Thanks to Daria Sorokina and Mona Sharma for their work on the project.

\bibliographystyle{plainnat}
\bibliography{references}

@inproceedings{tay2022dsi,
  title     = {Transformer Memory as a Differentiable Search Index},
  author    = {Tay, Yi and Tran, Vinh Q. and Dehghani, Mostafa and Ni, Jianmo and Bahri, Dara and Mehta, Harsh and Qin, Zhen and Hui, Kai and Zhao, Zhe and Gupta, Jai and Schuster, Tal and Cohen, William W. and Metzler, Donald},
  booktitle = {Advances in Neural Information Processing Systems (NeurIPS)},
  year      = {2022},
  url       = {https://arxiv.org/abs/2202.06991},
}

@inproceedings{decao2021genre,
  title     = {Autoregressive Entity Retrieval},
  author    = {De Cao, Nicola and Izacard, Gautier and Riedel, Sebastian and Petroni, Fabio},
  booktitle = {International Conference on Learning Representations (ICLR)},
  year      = {2021},
  url       = {https://arxiv.org/abs/2010.00904},
}

@inproceedings{covington2016youtube,
  title     = {Deep Neural Networks for {YouTube} Recommendations},
  author    = {Covington, Paul and Adams, Jay and Sargin, Emre},
  booktitle = {Proceedings of the 10th ACM Conference on Recommender Systems (RecSys)},
  pages     = {191--198},
  year      = {2016},
  doi       = {10.1145/2959100.2959190},
}

@inproceedings{yi2019twotower,
  title     = {Sampling-Bias-Corrected Neural Modeling for Large Corpus Item Recommendations},
  author    = {Yi, Xinyang and Yang, Ji and Hong, Lichan and Cheng, Derek Zhiyuan and Heldt, Lukasz and Kumthekar, Aditee and Zhao, Zhe and Wei, Li and Chi, Ed},
  booktitle = {Proceedings of the 13th ACM Conference on Recommender Systems (RecSys)},
  pages     = {269--277},
  year      = {2019},
  doi       = {10.1145/3298689.3346996},
}

@inproceedings{li2019mind,
  title     = {Multi-Interest Network with Dynamic Routing for Recommendation at {Tmall}},
  author    = {Li, Chao and Liu, Zhiyuan and Wu, Mengmeng and Xu, Yuchi and Zhao, Huan and Huang, Pipei and Kang, Guoliang and Chen, Qiwei and Li, Wei and Lee, Dik Lun},
  booktitle = {Proceedings of the 28th ACM International Conference on Information and Knowledge Management (CIKM)},
  pages     = {2615--2623},
  year      = {2019},
  doi       = {10.1145/3357384.3357814},
}

@inproceedings{zhai2024hstu,
  title     = {Actions Speak Louder than Words: Trillion-Parameter Sequential Transducers for Generative Recommendations},
  author    = {Zhai, Jiaqi and Liao, Lucy and Liu, Xing and Wang, Yueming and Li, Rui and Cao, Xuan and Gao, Leon and Gong, Zhaojie and Gu, Fangda and He, Michael and Lu, Yinghai and Shi, Yu},
  booktitle = {Proceedings of the 41st International Conference on Machine Learning (ICML)},
  year      = {2024},
  url       = {https://arxiv.org/abs/2402.17152},
}

@inproceedings{rajput2023tiger,
  title     = {Recommender Systems with Generative Retrieval},
  author    = {Rajput, Shashank and Mehta, Nikhil and Singh, Anima and Keshavan, Raghunandan and Vu, Trung and Heldt, Lukasz and Hong, Lichan and Tay, Yi and Tran, Vinh Q. and Samost, Jonah and Kula, Maciej and Chi, Ed H. and Sathiamoorthy, Maheswaran},
  booktitle = {Advances in Neural Information Processing Systems (NeurIPS)},
  year      = {2023},
  url       = {https://arxiv.org/abs/2305.05065},
}

@article{zhou2025openonerec,
  title   = {{OpenOneRec} Technical Report},
  author  = {Zhou, Guorui and Bao, Honghui and Huang, Jiaming and Deng, Jiaxin and Zhang, Jinghao and She, Junda and Cai, Kuo and Ren, Lejian and Ren, Lu and Luo, Qiang and Wang, Qianqian and Hu, Qigen and Zhang, Rongzhou and Tang, Ruiming and Wang, Shiyao and Li, Wuchao and Wu, Xiangyu and Luo, Xinchen and Wang, Xingmei and Hu, Yifei and Wu, Yunfan and Liu, Zhanyu and Zhang, Zhiyang and Zhang, Zixing and Chen, Bo and Wen, Bin and Ma, Chaoyi and Song, Chengru and Chu, Chenglong and Lian, Defu and Yang, Fan and Jiang, Feng and Cheng, Hongtao and Wang, Huanjie and Gai, Kun and Zheng, Pengfei and Wang, Qiang and Huang, Rui and Mao, Siyang and Gao, Tingting and Yuan, Wei and Wang, Yan and Zhou, Yang and Su, Yi and Cheng, Zexuan and Ling, Zhixin and Li, Ziming},
  journal = {arXiv preprint arXiv:2512.24762},
  year    = {2025},
  url     = {https://arxiv.org/abs/2512.24762},
}

@article{damico2026glide,
  title   = {Deploying Semantic {ID}-based Generative Retrieval for Large-Scale Podcast Discovery at {Spotify}},
  author  = {D'Amico, Edoardo and De Nadai, Marco and Chandar, Praveen and Vohra, Divita and Lin, Shawn and Lefarov, Max and Gigioli, Paul and Penha, Gustavo and Kopysitsky, Ilya and Senese, Ivo Joel and Mei, Darren and Fabbri, Francesco and Semerci, Oguz and Zhao, Yu and Tang, Vincent and St. Thomas, Brian and Ranieri, Alexandra and Smith, Matthew N. K. and Bernkopf, Aaron and Leung, Bryan and Fazelnia, Ghazal and VanMiddlesworth, Mark and Heath, Timothy Christopher and Skiden, Petter Pehrson and Wang, Alice Y. and Cole, Doug J. and Damianou, Andreas and Hristakeva, Maya and Wilbur, Reid and Chillara, Tarun and Radosavljevic, Vladan and Chitkara, Pooja and Adapa, Sainath and Elenter, Juan and Huber, Bernd and Wood, Jacqueline and Vedantam, Saaketh and Stypka, Jan and Ghael, Sandeep and Gould, Martin D. and Murgatroyd, David and Raimond, Yves and Lalmas, Mounia and Bennett, Paul N.},
  journal = {arXiv preprint arXiv:2603.17540},
  year    = {2026},
  url     = {https://arxiv.org/abs/2603.17540},
}

@inproceedings{ling2026crid,
  title     = {Beyond Semantic {IDs}: Encoding Business-Value Ranking into Document Identifiers for Generative Retrieval},
  author    = {Ling, Gui and Chen, Zhihong and Li, Yu and Xiong, Tong and Lin, Kunhai and Zhang, Kaixuan and Yan, Yuliang and Ou, Dan and Tang, Haihong and Zheng, Bo},
  booktitle = {Conference on Empirical Methods in Natural Language Processing (EMNLP): Industry Track},
  year      = {2026},
  url       = {https://arxiv.org/abs/2607.11392},
}

@article{robertson2009bm25,
  title   = {The Probabilistic Relevance Framework: {BM25} and Beyond},
  author  = {Robertson, Stephen and Zaragoza, Hugo},
  journal = {Foundations and Trends in Information Retrieval},
  volume  = {3},
  number  = {4},
  pages   = {333--389},
  year    = {2009},
  doi     = {10.1561/1500000019},
}

@article{furnas1987vocabulary,
  title   = {The Vocabulary Problem in Human-System Communication},
  author  = {Furnas, George W. and Landauer, Thomas K. and Gomez, Louis M. and Dumais, Susan T.},
  journal = {Communications of the ACM},
  volume  = {30},
  number  = {11},
  pages   = {964--971},
  year    = {1987},
  doi     = {10.1145/32206.32212},
}

@inproceedings{karpukhin2020dpr,
  title     = {Dense Passage Retrieval for Open-Domain Question Answering},
  author    = {Karpukhin, Vladimir and O{\u{g}}uz, Barlas and Min, Sewon and Lewis, Patrick and Wu, Ledell and Edunov, Sergey and Chen, Danqi and Yih, Wen-tau},
  booktitle = {Proceedings of the 2020 Conference on Empirical Methods in Natural Language Processing (EMNLP)},
  year      = {2020},
  url       = {https://arxiv.org/abs/2004.04906},
}

@inproceedings{nigam2019semantic,
  title     = {Semantic Product Search},
  author    = {Nigam, Priyanka and Song, Yiwei and Mohan, Vijai and Lakshman, Vihan and Ding, Weitian and Shingavi, Ankit and Teo, Choon Hui and Gu, Hao and Yin, Bing},
  booktitle = {Proceedings of the 25th ACM SIGKDD International Conference on Knowledge Discovery and Data Mining (KDD)},
  pages     = {2876--2885},
  year      = {2019},
  doi       = {10.1145/3292500.3330759},
}

@article{malkov2020hnsw,
  title   = {Efficient and Robust Approximate Nearest Neighbor Search Using Hierarchical Navigable Small World Graphs},
  author  = {Malkov, Yu. A. and Yashunin, D. A.},
  journal = {IEEE Transactions on Pattern Analysis and Machine Intelligence},
  volume  = {42},
  number  = {4},
  pages   = {824--836},
  year    = {2020},
  doi     = {10.1109/TPAMI.2018.2889473},
}

@inproceedings{lee2022rqvae,
  title     = {Autoregressive Image Generation using Residual Quantization},
  author    = {Lee, Doyup and Kim, Chiheon and Kim, Saehoon and Cho, Minsu and Han, Wook-Shin},
  booktitle = {IEEE/CVF Conference on Computer Vision and Pattern Recognition (CVPR)},
  year      = {2022},
  url       = {https://arxiv.org/abs/2203.01941},
}

@article{jegou2011pq,
  title   = {Product Quantization for Nearest Neighbor Search},
  author  = {J{\'e}gou, Herv{\'e} and Douze, Matthijs and Schmid, Cordelia},
  journal = {IEEE Transactions on Pattern Analysis and Machine Intelligence},
  volume  = {33},
  number  = {1},
  pages   = {117--128},
  year    = {2011},
}

@article{yang2025qwen3,
  title   = {{Qwen3} Technical Report},
  author  = {{Qwen Team}},
  journal = {arXiv preprint arXiv:2505.09388},
  year    = {2025},
  url     = {https://arxiv.org/abs/2505.09388},
}

@article{liquid2025lfm2,
  title   = {{LFM2} Technical Report},
  author  = {{Liquid AI}},
  journal = {arXiv preprint arXiv:2511.23404},
  year    = {2025},
  url     = {https://arxiv.org/abs/2511.23404},
}

@inproceedings{rafailov2023dpo,
  title     = {Direct Preference Optimization: Your Language Model is Secretly a Reward Model},
  author    = {Rafailov, Rafael and Sharma, Archit and Mitchell, Eric and Manning, Christopher D. and Ermon, Stefano and Finn, Chelsea},
  booktitle = {Advances in Neural Information Processing Systems (NeurIPS)},
  year      = {2023},
  url       = {https://arxiv.org/abs/2305.18290},
}

@article{shao2024deepseekmath,
  title   = {{DeepSeekMath}: Pushing the Limits of Mathematical Reasoning in Open Language Models},
  author  = {Shao, Zhihong and Wang, Peiyi and Zhu, Qihao and Xu, Runxin and Song, Junxiao and Bi, Xiao and Zhang, Haowei and Zhang, Mingchuan and Li, Y. K. and Wu, Y. and Guo, Daya},
  journal = {arXiv preprint arXiv:2402.03300},
  year    = {2024},
  url     = {https://arxiv.org/abs/2402.03300},
}

@article{hinton2015distillation,
  title   = {Distilling the Knowledge in a Neural Network},
  author  = {Hinton, Geoffrey and Vinyals, Oriol and Dean, Jeff},
  journal = {arXiv preprint arXiv:1503.02531},
  year    = {2015},
  url     = {https://arxiv.org/abs/1503.02531},
}

@article{ma2026mopd,
  title   = {{MOPD}: Multi-Teacher On-Policy Distillation for Capability Integration in {LLM} Post-Training},
  author  = {Ma, Wenhan and Wei, Jianyu and Zhao, Liang and Zhang, Hailin and Xiao, Bangjun and Li, Lei and Yang, Qibin and Gao, Bofei and Wang, Yudong and Li, Rang and Dong, Jinhao and Sui, Zhifang and Luo, Fuli},
  journal = {arXiv preprint arXiv:2606.30406},
  year    = {2026},
  url     = {https://arxiv.org/abs/2606.30406},
}

@inproceedings{ilharco2023taskarithmetic,
  title     = {Editing Models with Task Arithmetic},
  author    = {Ilharco, Gabriel and Ribeiro, Marco Tulio and Wortsman, Mitchell and Gururangan, Suchin and Schmidt, Ludwig and Hajishirzi, Hannaneh and Farhadi, Ali},
  booktitle = {International Conference on Learning Representations (ICLR)},
  year      = {2023},
  url       = {https://arxiv.org/abs/2212.04089},
}

@inproceedings{yadav2023ties,
  title     = {{TIES}-Merging: Resolving Interference When Merging Models},
  author    = {Yadav, Prateek and Tam, Derek and Choshen, Leshem and Raffel, Colin and Bansal, Mohit},
  booktitle = {Advances in Neural Information Processing Systems (NeurIPS)},
  year      = {2023},
  url       = {https://arxiv.org/abs/2306.01708},
}

@inproceedings{yu2024dare,
  title     = {Language Models are {Super Mario}: Absorbing Abilities from Homologous Models as a Free Lunch},
  author    = {Yu, Le and Yu, Bowen and Yu, Haiyang and Huang, Fei and Li, Yongbin},
  booktitle = {Proceedings of the 41st International Conference on Machine Learning (ICML)},
  year      = {2024},
  url       = {https://arxiv.org/abs/2311.03099},
}

@article{mccandlish2018largebatch,
  title   = {An Empirical Model of Large-Batch Training},
  author  = {McCandlish, Sam and Kaplan, Jared and Amodei, Dario and {OpenAI Dota Team}},
  journal = {arXiv preprint arXiv:1812.06162},
  year    = {2018},
  url     = {https://arxiv.org/abs/1812.06162},
}

@article{polyak1992averaging,
  title   = {Acceleration of Stochastic Approximation by Averaging},
  author  = {Polyak, Boris T. and Juditsky, Anatoli B.},
  journal = {SIAM Journal on Control and Optimization},
  volume  = {30},
  number  = {4},
  pages   = {838--855},
  year    = {1992},
  doi     = {10.1137/0330046},
}

@inproceedings{loshchilov2019adamw,
  title     = {Decoupled Weight Decay Regularization},
  author    = {Loshchilov, Ilya and Hutter, Frank},
  booktitle = {International Conference on Learning Representations (ICLR)},
  year      = {2019},
  url       = {https://arxiv.org/abs/1711.05101},
}

@article{hu2024minicpm,
  title   = {{MiniCPM}: Unveiling the Potential of Small Language Models with Scalable Training Strategies},
  author  = {Hu, Shengding and Tu, Yuge and Han, Xu and He, Chaoqun and Cui, Ganqu and Long, Xiang and Zheng, Zhi and Fang, Yewei and Huang, Yuxiang and Zhao, Weilin and Zhang, Xinrong and Thai, Zheng Leng and Zhang, Kaihuo and Wang, Chongyi and Yao, Yuan and Zhao, Chenyang and Zhou, Jie and Cai, Jie and Zhai, Zhongwu and Ding, Ning and Jia, Chao and Zeng, Guoyang and Li, Dahai and Liu, Zhiyuan and Sun, Maosong},
  journal = {arXiv preprint arXiv:2404.06395},
  year    = {2024},
  url     = {https://arxiv.org/abs/2404.06395},
}

@inproceedings{hagele2024scaling,
  title     = {Scaling Laws and Compute-Optimal Training Beyond Fixed Training Durations},
  author    = {H{\"a}gele, Alexander and Bakouch, Elie and Kosson, Atli and Ben Allal, Loubna and Von Werra, Leandro and Jaggi, Martin},
  booktitle = {Advances in Neural Information Processing Systems (NeurIPS)},
  year      = {2024},
  url       = {https://arxiv.org/abs/2405.18392},
}

@inproceedings{penedo2024fineweb,
  title     = {The {FineWeb} Datasets: Decanting the Web for the Finest Text Data at Scale},
  author    = {Penedo, Guilherme and Kydl{\'\i}{\v{c}}ek, Hynek and Ben Allal, Loubna and Lozhkov, Anton and Mitchell, Margaret and Raffel, Colin and Von Werra, Leandro and Wolf, Thomas},
  booktitle = {Advances in Neural Information Processing Systems (NeurIPS), Datasets and Benchmarks Track},
  year      = {2024},
  url       = {https://arxiv.org/abs/2406.17557},
}

@inproceedings{huang2025opencoder,
  title     = {{OpenCoder}: The Open Cookbook for Top-Tier Code Large Language Models},
  author    = {Huang, Siming and Cheng, Tianhao and Liu, Jason Klein and Xu, Weidi and Hao, Jiaran and Song, Liuyihan and Xu, Yang and Yang, Jian and Liu, Jiaheng and Zhang, Chenchen and Chai, Linzheng and Yuan, Ruifeng and Luo, Xianzhen and Wang, Qiufeng and Fan, YuanTao and Zhu, Qingfu and Zhang, Zhaoxiang and Gao, Yang and Fu, Jie and Liu, Qian and Li, Houyi and Zhang, Ge and Qi, Yuan and Xu, Yinghui and Chu, Wei and Wang, Zili},
  booktitle = {Proceedings of the 63rd Annual Meeting of the Association for Computational Linguistics (ACL)},
  year      = {2025},
  url       = {https://arxiv.org/abs/2411.04905},
}

@article{fujii2025swallowmath,
  title   = {Rewriting Pre-Training Data Boosts {LLM} Performance in Math and Code},
  author  = {Fujii, Kazuki and Tajima, Yukito and Mizuki, Sakae and Kawamura, Masaki and Shimada, Hinari and Shiotani, Taihei and Saito, Koshiro and Oi, Masanari and Nakamura, Taishi and Okamoto, Takumi and Ishida, Shigeki and Hattori, Kakeru and Ma, Youmi and Takamura, Hiroya and Yokota, Rio and Sakuma, Jun and Okazaki, Naoaki},
  journal = {arXiv preprint arXiv:2505.02881},
  year    = {2025},
  url     = {https://arxiv.org/abs/2505.02881},
}

@inproceedings{su2026static,
  title     = {Vectorizing the Trie: Efficient Constrained Decoding for {LLM}-based Generative Retrieval on Accelerators},
  author    = {Su, Zhengyang and Katsman, Isay and Wang, Yueqi and He, Ruining and Heldt, Lukasz and Keshavan, Raghunandan and Wang, Shao-Chuan and Yi, Xinyang and Gao, Mingyan and Dalal, Onkar and Hong, Lichan and Chi, Ed and Han, Ningren},
  booktitle = {Proceedings of the 32nd ACM SIGKDD Conference on Knowledge Discovery and Data Mining (KDD)},
  year      = {2026},
  pages     = {8009--8020},
  doi       = {10.1145/3770855.3818506},
  url       = {https://arxiv.org/abs/2602.22647},
}

@inproceedings{resnik1995information,
  title     = {Using Information Content to Evaluate Semantic Similarity in a Taxonomy},
  author    = {Resnik, Philip},
  booktitle = {Proceedings of the Fourteenth International Joint Conference on Artificial Intelligence (IJCAI)},
  year      = {1995},
  url       = {https://ijcai.org/proceedings/1995/59},
}

@article{jarvelin2002ndcg,
  title   = {Cumulated Gain-Based Evaluation of {IR} Techniques},
  author  = {J{\"a}rvelin, Kalervo and Kek{\"a}l{\"a}inen, Jaana},
  journal = {ACM Transactions on Information Systems},
  volume  = {20},
  number  = {4},
  pages   = {422--446},
  year    = {2002},
}

@article{dunnett1955,
  title   = {A Multiple Comparison Procedure for Comparing Several Treatments with a Control},
  author  = {Dunnett, Charles W.},
  journal = {Journal of the American Statistical Association},
  volume  = {50},
  number  = {272},
  pages   = {1096--1121},
  year    = {1955},
}

@inproceedings{hendrycks2021mmlu,
  title     = {Measuring Massive Multitask Language Understanding},
  author    = {Hendrycks, Dan and Burns, Collin and Basart, Steven and Zou, Andy and Mazeika, Mantas and Song, Dawn and Steinhardt, Jacob},
  booktitle = {International Conference on Learning Representations (ICLR)},
  year      = {2021},
  url       = {https://arxiv.org/abs/2009.03300},
}

@inproceedings{wang2018lambdaloss,
  title     = {The {LambdaLoss} Framework for Ranking Metric Optimization},
  author    = {Wang, Xuanhui and Li, Cheng and Golbandi, Nadav and Bendersky, Michael and Najork, Marc},
  booktitle = {Proceedings of the 27th ACM International Conference on Information and Knowledge Management (CIKM)},
  pages     = {1313--1322},
  year      = {2018},
  url       = {https://doi.org/10.1145/3269206.3271784},
}

@inproceedings{kim2016sequence,
  title     = {Sequence-Level Knowledge Distillation},
  author    = {Kim, Yoon and Rush, Alexander M.},
  booktitle = {Proceedings of the 2016 Conference on Empirical Methods in Natural Language Processing (EMNLP)},
  pages     = {1317--1327},
  year      = {2016},
  url       = {https://arxiv.org/abs/1606.07947},
}

\clearpage
\appendix
\section{Hyper-parameters}
\label{sec:appendix-hparams}

\Cref{tab:distillation-hparams} lists the settings of the models of \cref{sec:post-training-unification}.

\begin{table}[htbp]
  \caption{Settings of the models of \cref{sec:post-training-unification}, all started from the
    autoregressive CPT of \cref{tab:ar-mtp}. A node has eight B300 GPUs; all runs use AdamW
    ($\beta_1=0.9$, $\beta_2=0.999$), gradient clipping at norm 1.0 and seed 42. The two-stage
    distillation (\cref{sec:appendix-two-stage}) differs from distillation from the CPT in its student
    (joint SFT), its recommendation teacher (the 5,000-step LoRA expert), its rows (one \sid per row with
    format tokens, which the loss also covers) and its hardware (8 nodes, accumulation 4).}
  \label{tab:distillation-hparams}
  \centering\scriptsize\setlength{\tabcolsep}{3pt}
  \begin{tabular}{>{\raggedright\arraybackslash}p{0.72in} *{4}{>{\raggedright\arraybackslash}p{1.33in}}}
    \toprule
    & Joint SFT & Recommendation expert & Search expert & Distillation from the CPT \\
    \midrule
    Data & recommendation, search sessions, LLM-judged search (1:1:1) & recommendation & search sessions, LLM-judged search (1:1) & joint SFT's mix; prompt and three codewords \\
    Trained weights & all & LoRA (rank 64, $\alpha=128$, dropout 0.05, all linear layers) or all & all & all \\
    Steps & 5,000 & 5,000; teacher: 10,000 of 20,000 & 5,000 & 6,000 \\
    Batch per step & 256 packed sequences of 4,096 tokens & as joint SFT & as joint SFT & 2,048 rows, prompts up to 2,048 tokens \\
    Learning rate & $3\times10^{-5}$, 500 warmup steps, cosine to $6\times10^{-6}$ & LoRA: $10^{-4}$, 300 warmup steps (teacher: 1,200), cosine to $10^{-5}$; all: as joint SFT & as joint SFT & $10^{-5}$, 100 warmup steps, cosine to $5\times10^{-7}$ over the last 1,500 \\
    Weight decay & 0.1 & LoRA: 0; all: 0.1 & 0.1 & 0 \\
    Precision; averaging & bf16; none & bf16; none & bf16; none & fp32 master weights, bf16 autocast (teachers bf16); EMA, decay 0.998, published \\
    Loss & CE on completions & CE on completions & CE on completions & \cref{eq:distillation}, $\lambda=0.25$ \\
    Teachers & -- & -- & -- & LoRA rec.\ expert (10,000 steps); search expert \\
    Hardware & 1 node & 1 node (teacher: 4) & 1 node & 1 node; 8 rows per GPU, accumulation 32 (gold) or 4 and 64 (teacher prefix) \\
    \bottomrule
  \end{tabular}
\end{table}
\FloatBarrier

\section{Evaluation Protocols}
\label{sec:appendix-protocols}
\begin{table}[htbp]
  \caption{Evaluation protocols (\cref{sec:eval-protocols}). Each scores a set of search queries,
    and E3--E6 share one set. Judge: a personalized judge also reads the buyer's profile (\cref{sec:eval-ndcg}).
    Index: the products the \sidllm decodes over in search; E6 runs the whole search stack end to end
    (\cref{sec:eval-search-offline}). Users: the users scored for recommendation
    (\cref{sec:eval-rec-recall}); a dash marks a protocol without a recommendation evaluation.}
  \label{tab:eval-protocols}
  \centering\footnotesize
\begin{tabular}{l l l S[table-format=6] l}
\toprule
 & \multicolumn{2}{c}{Search} & \multicolumn{1}{c}{Recommendation} &  \\
\cmidrule(lr){2-3} \cmidrule(lr){4-4}
Protocol & Judge & Index & {Users} & Used in \\
\midrule
E1 & query only & top 50M & {--} & \cref{tab:rq-vae-vs-rq-kmeans} \\
E2 & query only & top 100M & {--} & \cref{tab:fit-sample,tab:codebook-sizes,tab:codebook-latency,tab:collision-dedup} \\
E3 & personalized & all eligible, 08-26 & 250000 & \cref{tab:product-attributes,tab:replay,tab:cross-task,tab:embedding-alignment,tab:mix-arms} \\
E4 & personalized & all eligible, 09-22 & 10000 & \cref{tab:anneal} \\
E5 & personalized & all eligible, 10-08 & 10000 & \cref{tab:ar-mtp,tab:post-training,tab:grpo,tab:backbones} \\
E6 & personalized & search stack, 10-26 & {--} & \cref{tab:search-stack} \\
\bottomrule
\end{tabular}

\end{table}
\FloatBarrier

\section{Teachers and distillation into joint SFT}
\label{sec:appendix-distillation}

\subsection{Choosing the teachers}
\label{sec:appendix-teachers}
For each task we trained a LoRA expert and a fully fine-tuned one with the same data and steps
(\cref{tab:distillation-hparams}), and kept as the teacher the one that scored higher on its own task
(\cref{tab:teacher-selection}). Search is scored on protocol E4, the only one on which both search
experts were scored. The two recipes also differ in learning rate, warmup, weight decay and dropout, so
the table compares tuned recipes rather than low-rank and full updates as such. The runs from the CPT
use the LoRA recommendation expert's checkpoint at step 10,000 of a 20,000-step schedule, chosen on the
same cohorts; it is not significantly better than the 5,000-step expert.

\begin{table}[htbp]
  \caption{Teacher selection: LoRA and fully fine-tuned experts for each task, with the same data and
    steps. First row: joint SFT's absolute scores; other rows: relative change (\%) from joint SFT with
    95\% paired-bootstrap intervals. Search: protocol E4 (\cref{tab:eval-protocols}), not comparable with
    E5. Bold: significant improvement; $\downarrow$: significant loss.}
  \label{tab:teacher-selection}
  \centering\footnotesize\setlength{\tabcolsep}{3pt}
  \IfFileExists{tables/teacher_selection.tex}{
\begin{tabular}{l c c c c}
\toprule
 & \multicolumn{2}{c}{Recommendation} & \multicolumn{2}{c}{Search (E4)} \\
\cmidrule(lr){2-3} \cmidrule(lr){4-5}
Model & {\recall{50}} & {\shortstack{Discovery\\\recall{50}}} & {\shortstack{Purchase\\\recall{10}}} & {\onedcg{10}} \\
\midrule
Joint SFT (absolute score) & \num{0.2411} & \num{0.0423} & \num{0.3439} & \num{0.3182} \\
\midrule
Rec. expert, LoRA & \bfseries \num[retain-explicit-plus]{+1.73}\% & \bfseries \num[retain-explicit-plus]{+7.81}\% & \num[retain-explicit-plus]{-1.91}\%\textsuperscript{$\downarrow$} & \num[retain-explicit-plus]{-2.30}\%\textsuperscript{$\downarrow$} \\
 & {\scriptsize [\num{0.94}, \num{2.51}]} & {\scriptsize [\num{4.45}, \num{11.05}]} & {\scriptsize [\num{-2.56}, \num{-1.20}]} & {\scriptsize [\num{-2.61}, \num{-2.00}]} \\
Rec. expert, full FT & \bfseries \num[retain-explicit-plus]{+1.37}\% & \bfseries \num[retain-explicit-plus]{+3.70}\% & \num[retain-explicit-plus]{-1.04}\%\textsuperscript{$\downarrow$} & \num[retain-explicit-plus]{-1.57}\%\textsuperscript{$\downarrow$} \\
 & {\scriptsize [\num{0.70}, \num{2.09}]} & {\scriptsize [\num{0.73}, \num{6.55}]} & {\scriptsize [\num{-1.66}, \num{-0.43}]} & {\scriptsize [\num{-1.83}, \num{-1.31}]} \\
\midrule
Search expert, LoRA & \num[retain-explicit-plus]{-20.83}\%\textsuperscript{$\downarrow$} & \num[retain-explicit-plus]{-32.69}\%\textsuperscript{$\downarrow$} & \num[retain-explicit-plus]{-0.59}\%\textsuperscript{$\downarrow$} & \num[retain-explicit-plus]{-0.03}\% \\
 & {\scriptsize [\num{-22.47}, \num{-19.18}]} & {\scriptsize [\num{-37.12}, \num{-28.19}]} & {\scriptsize [\num{-1.11}, \num{-0.08}]} & {\scriptsize [\num{-0.20}, \num{0.14}]} \\
Search expert, full FT & \num[retain-explicit-plus]{-20.19}\%\textsuperscript{$\downarrow$} & \num[retain-explicit-plus]{-32.12}\%\textsuperscript{$\downarrow$} & \num[retain-explicit-plus]{-0.30}\% & \num[retain-explicit-plus]{0.00}\% \\
 & {\scriptsize [\num{-21.82}, \num{-18.48}]} & {\scriptsize [\num{-36.54}, \num{-27.69}]} & {\scriptsize [\num{-0.80}, \num{0.23}]} & {\scriptsize [\num{-0.17}, \num{0.17}]} \\
\bottomrule
\end{tabular}
}{\todo{table pending}}
\end{table}

\subsection{The merge grid}
\label{sec:appendix-merging}
\begin{figure}[t]
  \centering
  \IfFileExists{figures/merge_frontier.pdf}{\includegraphics[width=\linewidth]{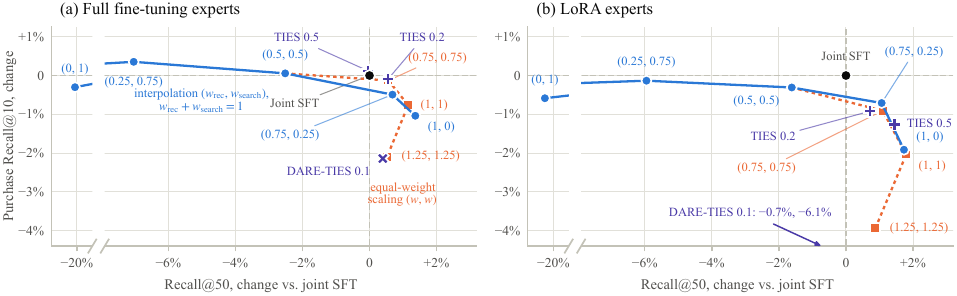}}{\todo{figure pending}}
  \caption{Weight merging slides along the experts' trade-off: merges of the fully fine-tuned (left)
    and LoRA (right) experts, labelled with their weights $(w_{\mathrm{rec}},w_{\mathrm{search}})$, as
    change (\%) from the fully fine-tuned joint SFT (the reference of both panels), with search on the
    earlier protocol E4. Solid:
    interpolation ($w_{\mathrm{rec}}+w_{\mathrm{search}}=1$); dotted: equal weights; TIES and DARE-TIES:
    weights (1,1) with the density shown (\cref{sec:appendix-merging}).}
  \label{fig:merge-frontier}
\end{figure}

The grid of \cref{fig:merge-frontier} merges experts of the same family on dense task vectors by
interpolation ($w_{\mathrm{rec}}+w_{\mathrm{search}}=1$), with equal weights of 0.75, 1 and 1.25, and at
(1,1) with TIES \citep{yadav2023ties} at densities 0.2 and 0.5 and with DARE \citep{yu2024dare} at
density 0.1 with TIES sign election. No tested merge is significantly better than joint SFT on
recommendation without being significantly worse on search.

\subsection{Distillation into joint SFT}
\label{sec:appendix-two-stage}
Before distilling into the CPT we distilled the 5,000-step experts into joint SFT, a two-stage recipe.
Our first runs followed MOPD (on-policy prefixes, no CE anchor, 256 rows per step, pure bf16) and left
the student where it started: at the starting weights the gradient noise scale
\citep{mccandlish2018largebatch} was about 1,100 rows for recommendation and 2,300 for search, against
about 85 and 170 such rows per step, and at a learning rate of $10^{-5}$ most AdamW updates were smaller
than half a bf16 unit in the last place and were rounded away. The runs of \cref{tab:kd-ce-ablation}
therefore use four times the batch, fp32 master weights and an exponential moving average of the weights.

\begin{figure}[htbp]
  \centering
  \IfFileExists{figures/last_codeword.pdf}{\includegraphics[width=\linewidth]{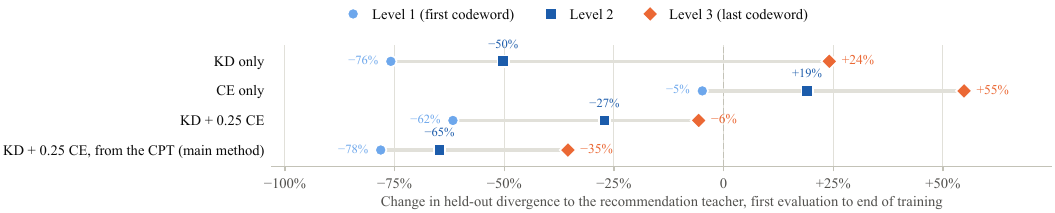}}{\todo{figure pending}}
  \caption{The last codeword is the hardest to distil: change in the held-out divergence per level from
    a run's first evaluation to its last (raw weights; runs into joint SFT: steps 200 to 2,600; from the
    CPT: 250 to 6,000). KD alone raises the third level, and the CE anchor reverses the rise. Top three
    rows: runs of \cref{tab:kd-ce-ablation}; bottom row: the main method, with its own teacher.}
  \label{fig:last-codeword}
\end{figure}

The first four rows of the table are matched, and \cref{fig:last-codeword} shows three of them per
level. Replacing the on-policy prefix by the gold prefix raised the recommendation point estimates but
lost more purchase recall, a loss that the CE anchor then removed; the two prefix runs were not compared
directly and their intervals overlap widely, so on-policy sampling bought nothing we could measure, and
the gold prefix, which needs no generation, became the default. The CE-only run has a different
coefficient from the combined run's, and none of these runs starts from the CPT.

The final two-stage run (last row) uses the settings of distillation from the CPT. It is the stronger of
the two recipes on search, with a significant gain in purchase \recall{10}, and its recommendation
recall is indistinguishable from its own teacher's. A no-teacher control at that scale, trained with the
CE term alone and evaluated after 3,000 steps, trailed distillation by 0.73\% [$+0.04$, $+1.48$]
\recall{50} in a paired comparison, and an interim second seed gained 0.98\% \recall{50} after 1,500
steps against 0.95\% for the first, the only measurement of training variance in this study.

\begin{table}[htbp]
  \caption{Distillation into joint SFT from the 5,000-step experts. The first four rows are matched
    (2.7M rows, 1,024 per step, constant learning rate $10^{-5}$, fp32 master weights, EMA decay 0.997).
    KD: the divergence of \cref{eq:distillation}, $\lambda=0$; CE only: the gold cross-entropy, weight 1,
    no teacher; KD + CE: $\lambda=0.25$. Last row: the final two-stage run, on 12.3M rows. Change (\%)
    from joint SFT with 95\% paired-bootstrap intervals; search: E5. Bold: significant improvement;
    $\downarrow$: significant loss. Last column: the third-level change of \cref{fig:last-codeword}, without an
    interval.}
  \label{tab:kd-ce-ablation}
  \centering\footnotesize\setlength{\tabcolsep}{3pt}
  \IfFileExists{tables/kd_ce_ablation.tex}{
\begin{tabular}{l c c c c c}
\toprule
 & \multicolumn{2}{c}{Recommendation} & \multicolumn{2}{c}{Search (E5)} & \multicolumn{1}{c}{Held-out} \\
\cmidrule(lr){2-3} \cmidrule(lr){4-5} \cmidrule(lr){6-6}
Model & {\recall{50}} & {\shortstack{Discovery\\\recall{50}}} & {\shortstack{Purchase\\\recall{10}}} & {\onedcg{10}} & {\shortstack{Level-3 KL\\change}} \\
\midrule
On-policy prefix, KD only & \bfseries \num[retain-explicit-plus]{+0.95}\% & \num[retain-explicit-plus]{+2.92}\% & \num[retain-explicit-plus]{-0.13}\% & \num[retain-explicit-plus]{-0.18}\% & \num[retain-explicit-plus]{+43.30}\% \\
 & {\scriptsize [\num{0.30}, \num{1.59}]} & {\scriptsize [\num{-0.11}, \num{5.95}]} & {\scriptsize [\num{-0.65}, \num{0.41}]} & {\scriptsize [\num{-0.38}, \num{0.00}]} &  \\
\midrule
KD only & \bfseries \num[retain-explicit-plus]{+1.05}\% & \bfseries \num[retain-explicit-plus]{+4.89}\% & \num[retain-explicit-plus]{-0.71}\%\textsuperscript{$\downarrow$} & \num[retain-explicit-plus]{-0.18}\% & \num[retain-explicit-plus]{+24.09}\% \\
 & {\scriptsize [\num{0.42}, \num{1.78}]} & {\scriptsize [\num{1.85}, \num{7.86}]} & {\scriptsize [\num{-1.26}, \num{-0.15}]} & {\scriptsize [\num{-0.37}, \num{0.01}]} &  \\
CE only & \num[retain-explicit-plus]{-0.37}\% & \num[retain-explicit-plus]{-1.83}\% & \num[retain-explicit-plus]{-1.03}\%\textsuperscript{$\downarrow$} & \num[retain-explicit-plus]{-0.66}\%\textsuperscript{$\downarrow$} & \num[retain-explicit-plus]{+54.81}\% \\
 & {\scriptsize [\num{-1.11}, \num{0.34}]} & {\scriptsize [\num{-4.67}, \num{1.14}]} & {\scriptsize [\num{-1.69}, \num{-0.38}]} & {\scriptsize [\num{-0.91}, \num{-0.42}]} &  \\
\(\mathrm{KD}+0.25\,\mathrm{CE}\) & \bfseries \num[retain-explicit-plus]{+1.12}\% & \bfseries \num[retain-explicit-plus]{+3.73}\% & \num[retain-explicit-plus]{+0.10}\% & \num[retain-explicit-plus]{-0.04}\% & \num[retain-explicit-plus]{-5.68}\% \\
 & {\scriptsize [\num{0.43}, \num{1.84}]} & {\scriptsize [\num{0.75}, \num{6.93}]} & {\scriptsize [\num{-0.44}, \num{0.66}]} & {\scriptsize [\num{-0.23}, \num{0.15}]} &  \\
\midrule
\(\mathrm{KD}+0.25\,\mathrm{CE}\), 12.3M rows (two-stage, final) & \bfseries \num[retain-explicit-plus]{+1.59}\% & \bfseries \num[retain-explicit-plus]{+8.01}\% & \bfseries \num[retain-explicit-plus]{+0.79}\% & \num[retain-explicit-plus]{-0.01}\% & \num[retain-explicit-plus]{-21.13}\% \\
 & {\scriptsize [\num{0.89}, \num{2.37}]} & {\scriptsize [\num{4.56}, \num{11.33}]} & {\scriptsize [\num{0.23}, \num{1.37}]} & {\scriptsize [\num{-0.20}, \num{0.19}]} &  \\
\bottomrule
\end{tabular}
}{\todo{table pending}}
\end{table}

\FloatBarrier

\begin{table}[htbp]
  \caption{Retargeting and discovery on the 9,990 users of the \recall{50} cohort: \recall{50} over
    purchased products already in the user's interaction history (retargeting; 5,295 users) and over the
    others (discovery; 6,281 users; not the 20,000-user cohort of \cref{tab:unification}). Discovery
    share: the part of the \recall{50} change from discovery purchases. New products: the share of the
    top 50 that the user never interacted with (absolute). First row: absolute baseline scores. Other
    recall cells: relative change (\%) from joint SFT with 95\% paired-bootstrap intervals. Bold:
    significant improvement; $\downarrow$: significant loss.}
  \label{tab:discovery-retargeting}
  \centering\footnotesize\setlength{\tabcolsep}{3pt}
  \IfFileExists{tables/discovery_retargeting.tex}{
\begin{tabular}{l c c c c}
\toprule
Model & {\shortstack{Retargeting\\\recall{50}}} & {\shortstack{Discovery\\\recall{50}}} & {\shortstack{Discovery share\\of the change}} & {\shortstack{New products\\in the top 50}} \\
\midrule
Joint SFT (absolute score) & \num{0.4747} & \num{0.0409} & -- & 72.9\% \\
\midrule
CPT & \num[retain-explicit-plus]{-18.45}\% & \num[retain-explicit-plus]{-32.36}\% & 14\% & 78.4\% \\
 & {\scriptsize [\num{-20.01}, \num{-16.91}]} & {\scriptsize [\num{-39.76}, \num{-24.28}]} &  &  \\
\midrule
Rec. expert (LoRA, 5k steps) & \bfseries \num[retain-explicit-plus]{+0.73}\% & \bfseries \num[retain-explicit-plus]{+11.88}\% & 68\% & 73.3\% \\
 & {\scriptsize [\num{0.10}, \num{1.34}]} & {\scriptsize [\num{5.61}, \num{18.68}]} &  &  \\
Rec. teacher (LoRA, 10k steps) & \num[retain-explicit-plus]{+0.76}\% & \bfseries \num[retain-explicit-plus]{+14.19}\% & 68\% & 74.6\% \\
 & {\scriptsize [\num{-0.04}, \num{1.55}]} & {\scriptsize [\num{6.38}, \num{22.69}]} &  &  \\
Search expert (full FT) & \num[retain-explicit-plus]{-19.37}\%\textsuperscript{$\downarrow$} & \num[retain-explicit-plus]{-32.46}\%\textsuperscript{$\downarrow$} & 13\% & 79.0\% \\
 & {\scriptsize [\num{-20.97}, \num{-17.83}]} & {\scriptsize [\num{-40.37}, \num{-24.54}]} &  &  \\
\midrule
Distillation, gold prefix & \num[retain-explicit-plus]{+0.53}\% & \bfseries \num[retain-explicit-plus]{+12.42}\% & 69\% & 73.9\% \\
 & {\scriptsize [\num{-0.24}, \num{1.27}]} & {\scriptsize [\num{5.19}, \num{20.58}]} &  &  \\
Distillation, teacher prefix & \num[retain-explicit-plus]{+0.27}\% & \bfseries \num[retain-explicit-plus]{+10.47}\% & 83\% & 74.1\% \\
 & {\scriptsize [\num{-0.53}, \num{0.99}]} & {\scriptsize [\num{3.22}, \num{18.57}]} &  &  \\
\bottomrule
\end{tabular}
}{\todo{table pending}}
\end{table}

\begin{table}[htbp]
  \caption{Baseline data mix of the LFM2-2.6B pre-training arms of \cref{sec:pretraining}: 30,000
    steps of 1,048,576 tokens. Columns as in \cref{tab:mixture}. The replay rows belong to the
    replay arm (\cref{tab:replay}), which gives general text 10\% of the loss tokens and scales the
    other streams to 90\%. Replay epochs are not computed.}
  \label{tab:baseline-mix}
  \centering\footnotesize
\begin{tabular}{l S[table-format=3.1] S[table-format=2.2] S[table-format=2.2] S[table-format=3.1] S[table-format=3.1]}
\toprule
Source & {Examples (M)} & {Epochs} & {Tokens (B)} & {Tokens (\%)} & {Loss tokens (\%)} \\
\midrule
\multicolumn{6}{l}{\itshape Product attributes} \\
\quad Product documents & 195.2 & 0.33 & 6.60 & 21.0 & 22.7 \\
\quad Top-GMV product documents & 200.3 & 8.00 & 5.99 & 19.0 & 23.3 \\
\addlinespace[3pt]
\multicolumn{6}{l}{\itshape Search} \\
\quad Search session examples & 37.2 & 1.00 & 12.23 & 38.9 & 5.3 \\
\quad LLM-judged examples & 33.4 & 2.00 & 0.76 & 2.4 & 3.2 \\
\quad LLM-judged personalized examples & 1.1 & 76.20 & 0.63 & 2.0 & 0.5 \\
\addlinespace[3pt]
\multicolumn{6}{l}{\itshape Recommendation} \\
\quad Interaction sequences & 8.4 & 0.02 & 2.88 & 9.2 & 22.5 \\
\quad New-product purchases & 12.0 & 0.05 & 1.36 & 4.3 & 13.5 \\
\quad New-merchant purchases & 13.0 & 0.06 & 0.77 & 2.4 & 6.7 \\
\quad New-category purchases & 3.5 & 0.02 & 0.25 & 0.8 & 2.3 \\
\midrule
Total & 504.0 & {--} & 31.46 & 100.0 & 100.0 \\
\midrule
\multicolumn{6}{l}{\itshape Replay (replay arm only)} \\
\quad FineWeb-Edu & 0.6 & {--} & 0.70 & 2.2 & 6.9 \\
\quad OpenCoder algorithmic & 1.0 & {--} & 0.19 & 0.6 & 1.9 \\
\quad SwallowMath-v2 & 0.1 & {--} & 0.12 & 0.4 & 1.2 \\
\bottomrule
\end{tabular}

\end{table}

\begin{table}[htbp]
  \caption{All LFM2-2.6B pre-training arms of \cref{sec:pretraining}, each one change from the
    baseline mix (\cref{tab:baseline-mix}). Evaluation protocol E3 (\cref{tab:eval-protocols}). A dash
    marks the task whose data the arm drops, or a task without an evaluation. Best value per column
    in bold.}
  \label{tab:mix-arms}
  \centering\footnotesize
\begin{tabular}{l S[table-format=1.4] S[table-format=1.4] S[table-format=1.4]}
\toprule
 & \multicolumn{1}{c}{Search relevance} & \multicolumn{1}{c}{Search purchases} & \multicolumn{1}{c}{Recommendation} \\
\cmidrule(lr){2-2} \cmidrule(lr){3-3} \cmidrule(lr){4-4}
Arm & {\onedcg{10}} & {\recall{10}} & {\recall{50}} \\
\midrule
Baseline & \bfseries 0.2876 & \bfseries 0.3336 & 0.2096 \\
\addlinespace[3pt]
\multicolumn{4}{l}{\itshape Product attributes} \\
\quad No product attribute data & 0.2818 & 0.3250 & \bfseries 0.2099 \\
\addlinespace[3pt]
\multicolumn{4}{l}{\itshape Search} \\
\quad No search data & {--} & {--} & 0.2087 \\
\quad Search session examples at 2.6\% of loss tokens & 0.2844 & 0.3179 & {--} \\
\addlinespace[3pt]
\multicolumn{4}{l}{\itshape Recommendation} \\
\quad No recommendation data & 0.2854 & 0.3263 & {--} \\
\quad Recommendation at 20\% of loss tokens & 0.2843 & 0.3322 & {--} \\
\addlinespace[3pt]
\multicolumn{4}{l}{\itshape Replay} \\
\quad General-text replay at 10\% of loss tokens & 0.2871 & 0.3276 & 0.2027 \\
\addlinespace[3pt]
\multicolumn{4}{l}{\itshape Training} \\
\quad No embedding alignment & 0.2860 & 0.3268 & 0.2067 \\
\bottomrule
\end{tabular}

\end{table}

\begin{figure}[t]
  \centering
  \includegraphics[width=\linewidth]{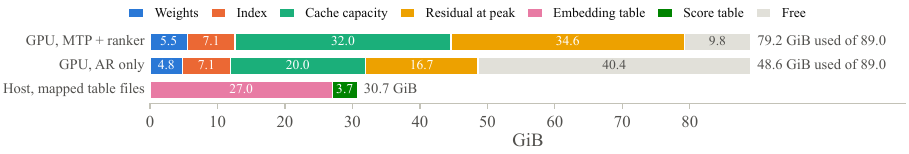}
  \caption{Memory accounting for AR only and MTP with the ranker on one B200 3g.90gb
    partition (89.0\,GiB), plus the ranker's host-mapped table files. Device totals are
    readings after a diagnostic run at 25 offered requests/s, with AR at beam 100 and MTP
    with the ranker at 250. Weight and index sizes are estimates; cache capacities are configured caps. Residual memory at
    peak is the measured device total minus these quantities, not an independently measured
    allocation. Host bars show table file sizes, not resident process memory.}
  \label{fig:inference-memory}
\end{figure}
\FloatBarrier

\section{Additional tokenizer experiments}
\label{sec:appendix-tokenizer}
This section gives the behavioral coherence, sampling, codebook, de-duplication and collision analyses supporting
\cref{sec:tokenizer}. Results use the evaluation protocol identified in each table; absolute
scores should only be compared within a table.

\subsection{Behavioral coherence of semantic IDs}
\label{sec:appendix-resnik}
We adapt Resnik similarity \citep{resnik1995information} to the tree formed by \sid prefixes,
measuring how specifically the tokenizers group distinct items purchased for the same query.
This is a retrospective analysis of the three tokenizers in \cref{tab:rq-vae-vs-rq-kmeans},
using behavioral data separate from the E1 downstream search evaluation.

We use search logs from August 1--26, 2026, retaining impressed, non-sponsored primary-variant
products from standard and agent search sessions with non-null product and session IDs.
Queries are lowercased, punctuation is replaced by spaces, and whitespace is collapsed and
trimmed; redacted or empty queries are excluded. An item is positive for a query if it has a
last-click-attributed purchase in at least one session for that query. Purchases are pooled
across sessions by normalized query.

We deduplicate listings by universal product ID (UPI). Each UPI uses one representative product,
shared across all three tokenizers, chosen among its impressed products in the May embedding
export. We select the product labeled with a purchase or high-quality click for the most
queries, breaking ties by the lowest product ID. UPIs without a representative are excluded,
retaining 68.6\% of purchased items. The common universe $U$ contains 41,913,701 distinct
impressed UPIs with a representative; each item contributes once to prefix frequencies.

Let $c(u)$ be item $u$'s \sid and $v(a,b)$ the longest common prefix of $c(a)$ and $c(b)$.
For prefix $v$ and two distinct items $a,b\in U$, define
\begin{equation}
  p(v) = \frac{\bigl|\{u\in U : c(u)\text{ has prefix }v\}\bigr|}{|U|},
  \qquad s(a,b) = -\log_2 p\bigl(v(a,b)\bigr).
  \label{eq:resnik-sid}
\end{equation}
The empty prefix has probability one, so pairs with different first codes score zero.
Pairs sharing their full \sid receive the information content of that full prefix. The
score therefore measures shared information in bits, rather than shared prefix length.

For each query $q$, let $S_q$ be its set of distinct purchased items after coverage filtering,
and let $\mathcal{P}_q$ be the set of unordered pairs of distinct items in $S_q$. We report
\begin{equation}
  R_q = \frac{1}{|\mathcal{P}_q|}\sum_{\{a,b\}\in\mathcal{P}_q}s(a,b),
  \qquad R = \frac{1}{|\mathcal{Q}|}\sum_{q\in\mathcal{Q}}R_q,
  \label{eq:resnik-query-mean}
\end{equation}
where $\mathcal{Q}$ contains the 12,799 queries with at least two retained purchased items.
Each query has equal weight. Repeating the analysis with high-quality clicks gives 5.93,
7.98 and 8.19 bits for \rqvae \mbox{1024-1024-1024-1024}, \rqkmeans \mbox{1024-1024-1024-1024} and \rqkmeans
\mbox{8192-8192-8192}, respectively, preserving the purchase-based ordering.

\subsection{Fitting sample}
We compare fitting on the top 25M products by GMV with fitting on a random 25M products from
the catalog. The top-GMV sample gives higher \onedcg{10} for both codebook shapes tested
(\cref{tab:fit-sample}).

\begin{table}[htbp]
  \caption{Codebook fit sample: the top 25M products by GMV vs.\ a random 25M, for an \rqkmeans
    tokenizer at two codebook shapes. We train downstream a 2.6B LLM for each of the methods, and evaluate performance in the generative retrieval task.
    Evaluation protocol E2 (\cref{tab:eval-protocols}). Better fit
    sample per codebook in bold.}
  \label{tab:fit-sample}
  \centering\footnotesize
\begin{tabular}{l S[table-format=1.4] S[table-format=1.4]}
\toprule
 & \multicolumn{2}{c}{\onedcg{10} by fit sample} \\
\cmidrule(lr){2-3}
Codebooks & {Top 25M by GMV} & {Random 25M} \\
\midrule
8192-8192-8192 & \bfseries 0.3464 & 0.3445 \\
32768-8192-8192 & \bfseries 0.3538 & 0.3496 \\
\bottomrule
\end{tabular}

\end{table}
\FloatBarrier

\subsection{Codebook size and balancing}
Increasing the codebook vocabulary generally improves downstream search relevance and the share
of unique \sids (\cref{tab:codebook-sizes}). Non-uniform codebooks perform similarly to uniform
ones with a similar vocabulary size. Balancing the last level raises uniqueness but lowers
\onedcg{10} at all but one tested shape, so we use standard $k$-means.

\begin{table}[htbp]
  \caption{Codebook shape and last-level balancing for an \rqkmeans tokenizer fit on the top 25M
    products by GMV. Evaluation protocol E2
    (\cref{tab:eval-protocols}). Best \onedcg{10} per column in bold.}
  \label{tab:codebook-sizes}
  \centering\footnotesize
\begin{tabular}{l S[table-format=5] S[table-format=2.1] S[table-format=1.4] S[table-format=2.1] S[table-format=1.4]}
\toprule
 &  & \multicolumn{2}{c}{Standard} & \multicolumn{2}{c}{Balanced last level} \\
\cmidrule(lr){3-4} \cmidrule(lr){5-6}
Codebooks & {SID tokens} & {Unique SIDs (\%)} & {\onedcg{10}} & {Unique SIDs (\%)} & {\onedcg{10}} \\
\midrule
1024-1024-1024 & 3072 & 12.9 & 0.2278 & 15.2 & 0.2253 \\
2048-2048-2048 & 6144 & 21.0 & 0.2376 & 24.1 & 0.2366 \\
4096-4096-4096 & 12288 & 27.5 & 0.2469 & 31.1 & 0.2458 \\
\midrule
8192-2048-512 & 10752 & 24.9 & 0.2466 & 27.3 & 0.2467 \\
8192-2048-1024 & 11264 & 27.0 & 0.2466 & 29.8 & 0.2450 \\
8192-4096-1024 & 13312 & 28.8 & \bfseries 0.2505 & 31.6 & \bfseries 0.2487 \\
\bottomrule
\end{tabular}

\end{table}
\FloatBarrier

\subsection{Serving latency across request rates}
\Cref{tab:codebook-latency-full} expands the target-load comparison of
\cref{tab:codebook-latency}. The larger codebook's p95 latency rises sharply beyond the target
load and reaches more than two seconds at 150 requests per second.

\begin{table}[htbp]
  \caption{Serving latency vs.\ codebook shape for the two \rqkmeans tokenizers of
    \cref{tab:fit-sample} fit on the top 25M products by GMV, each with its own LFM2-2.6B downstream LLM,
    on one NVIDIA B200 node. p95: the 95th-percentile latency at a fixed rate of requests per
    second. At the target load of 100 requests per second, the larger codebook adds 9.5\,ms
    (+23\%) to p95; at 150 it saturates. \onedcg{10}: evaluation protocol E2
    (\cref{tab:eval-protocols}). Better per column in bold.}
  \label{tab:codebook-latency-full}
  \centering\footnotesize
\begin{tabular}{l S[table-format=5] S[table-format=1.4] S[table-format=2.1] S[table-format=2.1] S[table-format=2.1] S[table-format=4.1]}
\toprule
 &  &  & \multicolumn{4}{c}{p95 latency (ms) at requests per second} \\
\cmidrule(lr){4-7}
Codebooks & {SID tokens} & {\onedcg{10}} & {50} & {100} & {125} & {150} \\
\midrule
8192-8192-8192 & 24576 & 0.3464 & \bfseries 29.6 & \bfseries 40.5 & \bfseries 51.6 & \bfseries 61.1 \\
32768-8192-8192 & 49152 & \bfseries 0.3538 & 31.7 & 50.0 & 93.2 & 2221.3 \\
\bottomrule
\end{tabular}

\end{table}
\FloatBarrier

\subsection{Heuristic de-duplication}
With the standard RQ-KMeans setup, we observe fairly low uniqueness (number of unique SIDs compared to the total number of products). Some
duplication is natural: the same product sold by several merchants has a different product id in each shop. Regardless, we attempt to overcome the duplication while maintaining or increasing downstream performance, and maintaining similar serving latency.
We explore two heuristic solutions: a random deduplication token and CRID \citep{ling2026crid}. In CRID, we choose a deduplication token based on the business-value ranking within each semantic cluster. For example, with CRID as token 4,
we first fit 3 codebooks with \rqkmeans and within each 3-token prefix assign a 4th token based on the GMV ranking of a product within that prefix. While these heuristics increase uniqueness, they decrease generative retrieval
metrics (\cref{tab:collision-dedup}) when we hold the codebook size constant. While adding a 4th CRID token improves downstream
retrieval performance slightly, it increases generation latency by 14\% on average, and up to 30\% at large beam widths and batch sizes.
Therefore, we don't adopt a heuristic deduplication strategy at tokenization time, but rather choose the simple \rqkmeans. 

\begin{table}[htbp]
  \caption{Heuristic de-duplication of colliding \sids on the production \mbox{8192-8192-8192}
    tokenizer. Evaluation protocol E2 (\cref{tab:eval-protocols}). Best \onedcg{10}
    per group in bold. Each group uses its own time-matched baseline.}
  \label{tab:collision-dedup}
  \centering\footnotesize
\begin{tabular}{l S[table-format=1] S[table-format=2.1] S[table-format=1.4]}
\toprule
Method & {SID length} & {Unique SIDs (\%)} & {\onedcg{10}} \\
\midrule
No de-duplication & 3 & 33.1 & 0.3204 \\
CRID as token 3 & 3 & 92.3 & 0.3180 \\
CRID as token 4 & 4 & 98.1 & \bfseries 0.3253 \\
\midrule
No de-duplication & 3 & 33.1 & \bfseries 0.3193 \\
Random token 4 & 4 & 96.8 & 0.3157 \\
\bottomrule
\end{tabular}

\end{table}
\FloatBarrier

\subsection{Cluster-size distribution}
\begin{figure}[htbp]
  \centering
  \includegraphics[width=\linewidth]{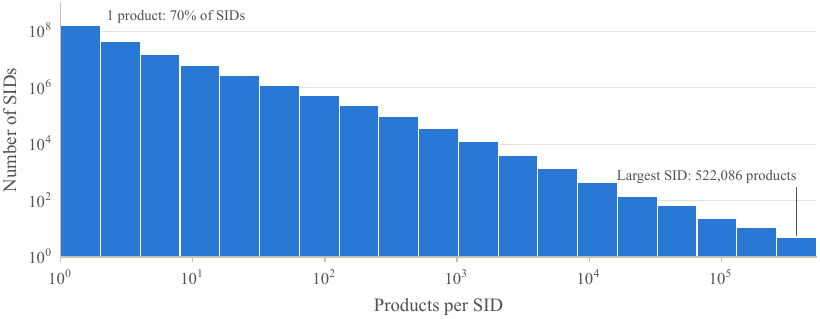}
  \caption{Products per \sid for the \mbox{8192-8192-8192} \rqkmeans tokenizer on a log-log plot. Counts in \cref{tab:sid-bucket-sizes}.}
  \label{fig:sid-bucket-sizes}
\end{figure}

Although the number of distinct \sids is low relative to the number of products, about 70\%
of occupied \sids contain exactly one product. The distribution is highly skewed
(\cref{fig:sid-bucket-sizes,tab:sid-bucket-sizes}). The largest clusters inspected contain
identical or near-identical long-tail listings, such as products whose titles differ only in
a part number (\cref{tab:largest-sids}); fewer than 0.3\% of the products in each of the three
largest clusters sold in the past year.

We also inspected reconstruction error at each level and code perplexity, but did not find
them predictive of downstream performance in our experiments. We therefore use retrieval
quality and latency to select the tokenizer (\cref{sec:tokenizer-model}).

\begin{table}[htbp]
  \caption{Products per \sid for the tokenizer of \cref{fig:sid-bucket-sizes}, by order of magnitude:
    the number of \sids in each size range, the products they hold, and those products as a share of
    all products.}
  \label{tab:sid-bucket-sizes}
  \centering\footnotesize
\begin{tabular}{l S[table-format=9] S[table-format=9] S[table-format=3.1]}
\toprule
Products per SID & {SIDs} & {Products} & {Share of products (\%)} \\
\midrule
1 & 162505415 & 162505415 & 22.8 \\
2--9 & 60670564 & 196267628 & 27.5 \\
10--99 & 7974538 & 185733270 & 26.0 \\
100--999 & 512471 & 115602794 & 16.2 \\
1,000--9,999 & 18436 & 37966316 & 5.3 \\
10,000--99,999 & 513 & 11180570 & 1.6 \\
100,000--522,086 & 22 & 4700480 & 0.7 \\
\midrule
All & 231681959 & 713956473 & 100.0 \\
\bottomrule
\end{tabular}

\end{table}

\begin{table}[htbp]
  \caption{The three largest \sids of \cref{fig:sid-bucket-sizes}, by title pattern: titles that are
    identical, ignoring case, once every word containing a digit (such as a part number or a size) is
    masked. Each row shows one title of the pattern, with the brand name replaced by [brand], and the
    number of products in the pattern.}
  \label{tab:largest-sids}
  \centering\footnotesize
\begin{tabular}{l S[table-format=6]}
\toprule
Example title & {Products} \\
\midrule
\multicolumn{2}{l}{\itshape Largest \sid: 522,086 products, 99.99\% from one shop} \\
\quad [brand] SAU Series: Double-axis Compound Stroke Adjustable Cylinder - SAUJ63X250-20S & 520819 \\
\quad [brand] SAI Series: Double Axis Compound Stroke Adjustable Cylinder - SAIJ32X175-10H & 831 \\
\quad [brand] SAU Series: Standard Compound Cylinder - SAU32X70S & 349 \\
\quad 3 other title patterns & 87 \\
\addlinespace[3pt]
\multicolumn{2}{l}{\itshape Second-largest \sid: 357,508 products, all from one shop} \\
\quad [brand] 1200 Boat Cover & 327270 \\
\quad [brand] 600 Pontoon Cover & 29876 \\
\quad [brand] 1200 Jet Ski Cover & 362 \\
\addlinespace[3pt]
\multicolumn{2}{l}{\itshape Third-largest \sid: 351,446 products, 99.94\% from one shop} \\
\quad [brand] SE: Double Acting Standard Air Cylinder - SE80X75S & 232858 \\
\quad [brand] SC series: double-shaft double-acting cylinder - SCD40X120 & 47291 \\
\quad [brand] SC series: Standard double-acting cylinder - SC50X1300STC & 40952 \\
\quad 115 other title patterns & 30343 \\
\bottomrule
\end{tabular}

\end{table}
\FloatBarrier

\section{Additional task-unification analysis}
\label{sec:appendix-unification-details}
This section provides evaluation details and supporting experiments for
\cref{sec:post-training-unification}. Teacher selection, the merge grid and the earlier two-stage
recipe are described in \cref{sec:appendix-distillation}.

\subsection{Evaluation cohorts}
\label{sec:appendix-unification-eval}
All changes are relative to joint SFT, $100\,(m/m_{\mathrm{joint}}-1)$ for a metric $m$. Joint SFT
samples recommendation examples and two search streams in a 1:1:1 ratio
(\cref{tab:distillation-hparams}). Recommendation \recall{50} is evaluated over 9,990 held-out
purchasers using 50 beam-decoded \sids (\cref{sec:eval-rec-recall}). Discovery \recall{50} restricts
targets to products absent from the user's interaction history and uses a separate cohort of 20,000
users (\cref{sec:pretraining-rec-data}). Search uses the purchase \recall{10} of 25,000 converted
sessions and the \onedcg{10} of 6,000 queries under protocol E5
(\cref{sec:eval-purchase-recall,sec:eval-ndcg,tab:eval-protocols}).

We compute 95\% paired-bootstrap intervals with 2,000 resamples. A change is significant when its
interval excludes zero; an undetected change does not establish equivalence. Each model was trained
once, and the development cohorts and queries also guided teacher, checkpoint and recipe selection,
so the intervals are nominal.

\subsection{Weight-merging results}
\label{sec:unification-merging}
We explore combining expert weights through task arithmetic, interpolation and sparse merging
\citep{ilharco2023taskarithmetic,yadav2023ties,yu2024dare}. The grid is specified in
\cref{sec:appendix-merging}. None of the tested merges significantly improves recommendation
without a significant search loss. The figure uses the earlier search protocol E4; the main results
table uses E5.

\subsection{Distillation from the CPT}
The student trains on 12.3M rows with a warmup-stable-decay schedule
\citep{hu2024minicpm,hagele2024scaling}, a large batch and fp32 master weights
(\cref{sec:appendix-two-stage}), and we evaluate an exponential moving average of its weights
\citep{polyak1992averaging}.

\subsubsection{Objective and validation diagnostics}
Our recipe adapts multi-teacher on-policy distillation (MOPD; \citealt{ma2026mopd}), retaining
per-task routing and exact per-token reverse KL. The student starts from the CPT, the divergence is
evaluated on gold prefixes rather than sampled student prefixes, as in word-level distillation
\citep{kim2016sequence}, and a cross-entropy term anchors the student to the gold item.
The \emph{held-out divergence} is the student's exact per-level
$\mathrm{KL}(P_\theta\,\|\,T)$ from its recommendation teacher, teacher-forced on gold items in fixed
validation prompts.

The case for the teachers and for the anchor comes from matched runs of an earlier recipe that
distilled the 5,000-step experts into joint SFT (\cref{tab:kd-ce-ablation}). CE alone, in effect
continued joint SFT, gained nothing detectable on recommendation and lost search; distillation alone
gained on recommendation but lost purchase \recall{10} significantly; both together kept the gain with
no detected search change. Distillation alone lowered the held-out divergence at the first two levels
and raised it at the third, and the anchor reversed the rise (\cref{fig:last-codeword}); our untested
reading is that the shared update fits the broader levels at the expense of the more constrained third,
which the hard label holds in place. $\lambda=0.25$ equalises the two terms' initial gradient norms and
was not tuned. The CPT recipe has no matched KD-only or CE-only run, so these runs motivate its anchor
rather than establish that it needs one.

\subsubsection{Gold or teacher prefix}
\label{sec:unification-prefix}
\begin{figure}[t]
  \centering
  \IfFileExists{figures/prefix_sampling.pdf}{\includegraphics[width=\linewidth]{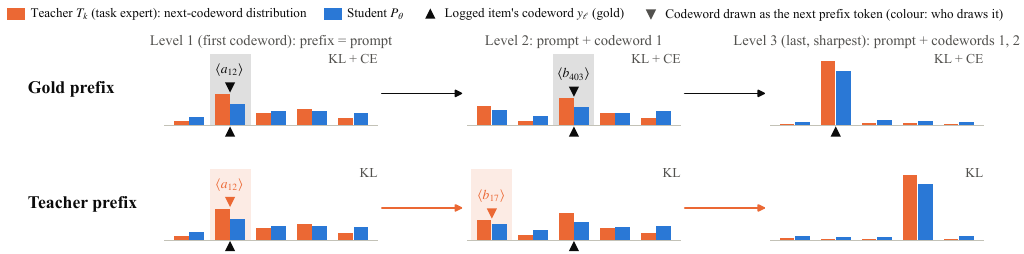}}{\todo{figure pending}}
  \caption{Where the divergence of \cref{eq:distillation} is taken, for one row whose gold item is
    $\langle a_{12}\rangle\langle b_{403}\rangle\langle c_{77}\rangle$ (schematic; illustrative
    distributions). Gold prefix (top): every level is teacher-forced on the gold item, so KL and CE fall
    on the same tokens. Teacher prefix (bottom): the teacher's sampled \sid, drawn once per row and step and not
    differentiated, sets the later levels' prefix and can leave the gold path ($\langle b_{17}\rangle$), while CE stays on the
    gold item; in training it is the gold item for 9\% of rows and shares its first codeword for 45\%.}
  \label{fig:prefix-sampling}
\end{figure}

Beam search ranks 50 paths, but the gold prefix trains only the purchased item's. A prefix drawn from the
teacher, one \sid per row and step at temperature 1 with each level restricted to its own codewords,
also covers the other branches in proportion to the teacher's probability, while CE stays on the gold
item (\cref{fig:prefix-sampling}). In a second run from the CPT, with the same teachers, rows per step and
schedule, it fits the teacher more closely (held-out divergence 0.060 against 0.072 at step 6,000) but
retrieves no better: paired with the gold prefix, \recall{50} changes by $-0.46\%$ [$-0.98$, $+0.04$],
and no direct contrast on either task is significant. Its draw is the gold item in only 9\% of rows, so
most third-level targets lie off the purchased path. We keep the gold prefix and read the held-out
divergence as a diagnostic, not a criterion for choosing models.

\subsection{Retargeting and discovery}
\label{sec:appendix-unification-discovery}
\recall{50} is dominated by retargeting: 90\% of joint SFT's hits in its top 50 are products already in
the buyer's interaction history. On the same users, recommendation training moves discovery recall by
10.5--14.2\% but retargeting recall by only 0.3--0.8\%, so discovery purchases, a tenth of the hits,
supply 68--83\% of each model's gain (\cref{tab:discovery-retargeting}). The CPT and the search expert
put more products new to the buyer in their top 50 (78--79\% against 73\%) yet lose both kinds of recall:
recommending unseen products is not discovering the ones the buyer will buy. Panel (c) of
\cref{fig:unification-frontier} therefore separates the routes most clearly, although discovery recall
stays small, 4.1\% for joint SFT.
\FloatBarrier

\section{Search post-training details}
\label{sec:appendix-search-post-training}
This section gives the data construction, objectives and training settings for the search experiments
in \cref{sec:post-training-search}. Each method starts independently from the same CPT+MTP checkpoint.
The judge-derived scores below are training signals; the primary evaluation metric is
\onedcg{10} under protocol E5.

\subsection{Offline data construction}
\label{sec:appendix-post-training-data}
The offline data consist of search sessions with a query, an optional buyer profile and candidate
products scored by a fine-tuned Qwen3.5-2B relevance model. Candidates come from search engine
results pages (SERPs) and, where available, up to 200 products generated by the CPT+MTP model per
session. We remove unjudgeable candidates and non-finite scores, and deduplicate complete \sids
within each session by keeping the highest score.

The corpus contains approximately 1.8M session records. Positive filtering retains an estimated
1.38M records for CE with SERP targets and 1.49M for CE+DPO with SERP and generated targets.
CE+DPO reserves approximately 9,000 sessions to monitor preference losses. We compare retrieval
performance using protocol E5 (\cref{tab:eval-protocols}).

\subsection{Supervised fine-tuning details}
\label{sec:appendix-post-training-ce}
For soft-target CE, we retain candidates with judge score at least 0.71. We choose this value,
approximately the mean judge score plus 0.75 standard deviations, to retain enough positive
products in the training data. If candidate $i$ has score $r_i$ and three-codeword \sid $y_i$,
its weight and the target distribution at level $\ell$ are
\begin{equation}
  q_i = \frac{\exp(r_i/\tau)}{\sum_j \exp(r_j/\tau)},
  \qquad Q_\ell(t) = \sum_i q_i\,\mathbf{1}[y_{i,\ell}=t],
  \qquad \tau = 0.05.
\end{equation}
The codeword loss is $-\sum_{\ell=1}^{3}\sum_t Q_\ell(t)\log P_{\theta,\ell}(t\mid x)$,
where $x$ is the query and profile prompt and $P_{\theta,\ell}$ is a full-vocabulary probability.
For fixed targets, each level's CE equals forward KL divergence from $Q_\ell$ plus its entropy.
The objective matches the judge-derived codeword marginals. We add hard-label CE on
completion-format tokens and divide the summed loss by the number of supervised tokens. Prompts
carry no loss. The MTP readouts predict the three codewords in parallel without seeing target
codewords in their inputs.

The offline prompt includes a new relevance token to signal that the model should generate
products with high judged relevance. This distinguishes the new output objective from predicting
observed purchases. We initialise the new token's embedding from the existing purchase token.
We pack examples into 4,096-token sequences. Both offline methods use AdamW, a peak learning rate of $10^{-5}$, 100 warmup steps,
cosine decay, weight decay 0.1, gradient clipping at 1.0, bf16 and seed 42, on eight B300 GPUs.
\Cref{tab:post-training-recipes} gives the schedule lengths and batch units. Each packed sequence
can contain several queries. We train CE for 1,600 steps without a validation set. Both offline
schedules use fixed step budgets without early stopping.

\begin{table}[htbp]
  \caption{Post-training schedules. All runs use eight trainer GPUs and a 100-step learning-rate
    warmup. Offline batch sizes count packed sequences per GPU; GRPO counts candidate completions,
    32 per query, and uses one additional B300 GPU for the judge. The constant GRPO learning rate
    follows warmup. Batch sizes across these units are not directly comparable.}
  \label{tab:post-training-recipes}
  \centering\footnotesize
\begin{tabular}{l S[table-format=5] S[table-format=2] l l}
\toprule
Method & {Steps} & {Batch/GPU} & Batch unit & LR schedule \\
\midrule
Soft-target CE & 1600 & 8 & Packed sequences & Cosine \\
CE+DPO & 1250 & 8 & Packed sequences & Cosine \\
GRPO & 50000 & 32 & Completions & Constant \\
\bottomrule
\end{tabular}

\end{table}

\subsection{Offline CE+DPO details}
\label{sec:appendix-post-training-dpo}
CE+DPO adds a reference-relative preference loss to positive-target CE \citep{rafailov2023dpo}.
After deduplication, positives score above 0.71 and negatives below it. We omit scores of exactly
0.71 and queries without positives; positive-only queries retain CE. Weights and confidences
depend on distance from the threshold:
\begin{equation}
  d_i=|r_i-0.71|,\qquad
  q_i^{\pm}=\frac{\exp(d_i/\tau)}{\sum_{j\in\mathcal{Y}^{\pm}}\exp(d_j/\tau)},\qquad
  c^{\pm}=\sum_{i\in\mathcal{Y}^{\pm}}q_i^{\pm}\bigl(1-\exp(-d_i/\tau)\bigr).
\end{equation}
Let $s_\theta(y)=\sum_{\ell=1}^{3}\log P_{\theta,\ell}(y_\ell\mid x)$ and let
$s_{\mathrm{ref}}$ be the same score under the fixed CPT+MTP initialisation. For a positive and
negative pair, define
\begin{equation}
  \Delta_\theta(y^+,y^-)
  =s_\theta(y^+)-s_\theta(y^-)-s_{\mathrm{ref}}(y^+)+s_{\mathrm{ref}}(y^-).
\end{equation}
The per-query codeword terms are
\begin{align}
  \mathcal{L}_{\mathrm{CE}} &=-\frac{c^+}{3}\,\E_{q^+}[s_\theta(y^+)],\\
  \mathcal{L}_{\mathrm{DPO}} &=c^+c^-\,\E_{q^+q^-}
    \left[\log\bigl(1+\exp(-\beta\Delta_\theta(y^+,y^-))\bigr)\right],\qquad \beta=0.1.
\end{align}
For each query, we compute the preference loss over its positive and negative candidates,
weighted by $q^+q^-$. We average these terms over queries, not confidence mass, and add a
separately token-normalised format CE term. We weight CE, DPO and format CE equally, with
$\tau=0.05$, and train for 1,250 steps. We also train a SERP-only variant and an all-source variant
with the DPO coefficient doubled, each for 1,250 steps. We monitor preference losses on the
held-out sessions during training.

\subsection{GRPO details}
\label{sec:appendix-post-training-grpo}
For online optimisation, we use an asynchronous variant of GRPO \citep{shao2024deepseekmath}. For each query,
an MTP trie traversal with beam width 256 proposes the top 32 three-token \sids, renormalising
probabilities over allowed successors. These are ranked proposals rather than independent policy
samples. For each \sid, we look up product information and score the first product returned with
the query and buyer profile, using our Qwen3.5-2B training judge.

Rewards are centred by the query group's mean and scaled by a component-level exponential moving
standard deviation, with decay 0.995, a 200-batch warmup and a 0.01 floor. We assign unscored candidates the group mean. The policy objective uses
token-level probability ratios, clipping at 0.2 and a KL coefficient of
0.04 against the fixed initial policy, with one update per generated batch and a loss averaged
over tokens. We refresh generation weights every ten optimiser steps; generation and reward
queues each hold up to eight batches. This procedure differs from the original GRPO in its ranked proposals,
reward scaling and asynchronous updates.

We train for 50,000 steps with a global batch of 256 completions:
eight queries with 32 proposals each. The training batches therefore contain 400,000 query
presentations and 12.8M completions in total. We use AdamW at $10^{-5}$ after warmup, gradient
clipping at 1.0, bf16 and seed 42.

\subsection{Interpreting the method comparisons}
\Cref{tab:post-training} compares generator-only results under protocol E5. Our primary metric is \onedcg{10}.
As defined in \cref{sec:eval-ndcg}, this metric normalises by a constant theoretical maximum.
We treat an absolute increase of at least 0.002 as a practically meaningful improvement; this is a decision threshold,
not a statistical significance test.

GRPO achieves the highest score among the three methods, exceeding
all-source CE+DPO with equal coefficients by 0.0021 and soft-target CE by 0.0040. All-source
CE+DPO exceeds CE by 0.0020, close to the practical threshold, but adds generated candidates
and changes the training schedule and validation split. It therefore does not isolate the effect
of the preference loss. The closest SERP-only CE+DPO comparison gains 0.0007 over CE, below the
threshold. Doubling the all-source DPO coefficient gives no practically meaningful improvement.

\subsection{Evaluation and attribution limits}
\label{sec:appendix-post-training-controls}
A matched all-source CE baseline with the CE+DPO corpus, 1,250-step schedule and inference setup
is still needed to attribute any gain to DPO. The models share evaluation query sets and a \sid
trie; common decoding and serving settings would further control the comparison. Other remaining
work includes per-query uncertainty estimates, repeated runs with different seeds, an audit of
training-data overlap with E5, and verification of the independence of the training and evaluation
judges. These results describe performance on the development benchmark. Complete reproducibility
also requires archiving the GRPO source revision or training-image digest and its mapping to the
evaluated model. Applying GRPO after the strongest offline method remains a subsequent experiment.

\subsection{Additional GRPO comparison}
The following comparison retains the results with the lightweight ranker. The ranker backend changes
between the two configurations, so the pair does not isolate the effect of GRPO.

\begin{table}[htbp]
  \caption{GRPO on the multi-token prediction model of \cref{tab:ar-mtp}, without and with the ranker.
    Evaluation protocol E5 (\cref{tab:eval-protocols}). Better of each pair in
    bold. The ranker weights are shared, but the baseline ranker uses the HF backend and the GRPO
    ranker uses Liquid; that pair does not isolate the effect of GRPO.}
  \label{tab:grpo}
  \centering\footnotesize
\begin{tabular}{l S[table-format=1.4] S[table-format=1.4]}
\toprule
 & \multicolumn{1}{c}{Search relevance} & \multicolumn{1}{c}{Search purchases} \\
\cmidrule(lr){2-2} \cmidrule(lr){3-3}
Arm & {\onedcg{10}} & {\recall{10}} \\
\midrule
\multicolumn{3}{l}{\itshape Multi-token prediction} \\
\quad Without GRPO & 0.2811 & \bfseries 0.2840 \\
\quad With GRPO & \bfseries 0.2970 & 0.2782 \\
\addlinespace[3pt]
\multicolumn{3}{l}{\itshape Multi-token prediction + ranker} \\
\quad Without GRPO & 0.3365 & \bfseries 0.3450 \\
\quad With GRPO & \bfseries 0.3453 & 0.3328 \\
\bottomrule
\end{tabular}

\end{table}
\FloatBarrier

\end{document}